\documentclass[aps,prd,11pt,tightenlines,superscriptaddress,nofootinbib,preprintnumbers,notitlepage]{revtex4-1}
 
\pdfoutput=1
\usepackage{color,amsmath,amssymb,multirow,hyperref,graphicx,tabularx,slashed}

\def\0#1#2{\frac{#1}{#2}}

\newcommand\one{\leavevmode\hbox{\small1\normalsize\kern-.33em1}}

\newcommand{\lag}{\mathcal{L}}

\newcommand{\qqquad}{\qquad \qquad}
\newcommand{\qqqquad}{\qquad \qquad \qquad}

\newcommand{\gev}{\text{GeV}}
\newcommand{\tev}{\text{TeV}}

\newcommand{\iab}{\text{ab}^{-1}}

\newcommand{\eg}{\textsl{e.g.}\;}
\newcommand{\ie}{\textsl{i.e.}\;}

\newcommand{\Mats}{\sum_{n\in\mathbb Z}}

\graphicspath{{./figs/}}

\begin{document}
 
\title{Probing Baryogenesis through the Higgs Self-Coupling}

\author{M.~Reichert}

\affiliation{Institut f\"ur Theoretische Physik, Universit\"at Heidelberg, Germany}
\author{A.~Eichhorn}

\affiliation{Institut f\"ur Theoretische Physik, Universit\"at Heidelberg, Germany}

\author{H.~Gies}

\affiliation{Theoretisch-Physikalisches Institut, Abbe Center of Photonics, Friedrich-Schiller-Universit\"at Jena, Germany}
\affiliation{Helmholtz Institute Jena, Germany}

\author{J.~M.~Pawlowski}

\affiliation{Institut f\"ur Theoretische Physik, Universit\"at Heidelberg, Germany}
\affiliation{Extreme Matter Institute, GSI, Darmstadt, Germany}

\author{T.~Plehn}

\affiliation{Institut f\"ur Theoretische Physik, Universit\"at Heidelberg, Germany}

\author{M.~M.~Scherer}

\affiliation{Institut f\"ur Theoretische Physik, Universit\"at zu K\"oln, Germany}

\begin{abstract}
  The link between a modified Higgs self-coupling and the strong
  first-order phase transition necessary for baryogenesis is well
  explored for polynomial extensions of the Higgs potential. We
  broaden this argument beyond leading polynomial expansions of the
  Higgs potential to higher polynomial terms and to non-polynomial
  Higgs potentials. For our quantitative analysis we resort to the
  functional renormalization group, which allows us to evolve the full
  Higgs potential to higher scales and finite temperature. In all
  cases we find that a strong first-order phase transition manifests
  itself in an enhancement of the Higgs self-coupling by at least
  50\%, implying that such modified Higgs potentials should be
  accessible at the LHC.
\end{abstract}

\maketitle

\medskip 
\medskip 

\tableofcontents 

\clearpage

\section{Introduction}
\label{sec:introduction}

The existence of a scalar Higgs potential is the most fundamental
insight from the LHC to date. It is based on the observation of a
likely fundamental Higgs scalar in combination with measurements of
the massive electroweak bosons, fixing the infrared theory and its
model parameters after electroweak symmetry breaking to high
precision. The one remaining parameter is the Higgs self-coupling and
its relation to the Higgs mass, defining a standard benchmark
measurement for current and future colliders.  
This in itself very interesting measurement may also be related 
to more fundamental physics questions. A prime candidate
for such a question is electroweak baryogenesis, specifically the
nature of the electroweak phase transition. 

For the single Higgs boson of the renormalizable Standard Model we can
test the electroweak phase transition through the Higgs mass.  Here,
electroweak baryogenesis~\cite{ew_phase,review_ew} requires a Higgs
mass well below the observed value of
125~GeV~\cite{misha_higgs,ew_higgs,Shaposhnikov:1991cu}. Only then will the electroweak
phase transition be strongly first order. 
If we consider the Standard
Model an effective field theory (EFT), a sizable dimension-6 contribution to the
Higgs potential, $(\phi^\dag \phi)^3/\Lambda^2$, is known to
circumvent this bound~\cite{eft1,eft2,Noble,christophe_geraldine}. 
In principle, this scenario
can be tested through a measurement of the Higgs self-coupling at
colliders~\cite{uli1,eft2,Noble,christophe_geraldine}. The problem with this link is that the 
new-physics scale
required by a first-order phase transition is typically
not large, $\Lambda \gtrsim v = 246$~GeV. If LHC data should indeed
point to a dimension-6 Lagrangian with a low new-physics scale, we
will see this in many other channels long before we will actually
measure the Higgs self-coupling~\cite{legacy}.
As a matter of fact, a global analysis of the effective Higgs
Lagrangian including $(\phi^\dag \phi)^3/\Lambda^2$ might
never probe the required values of the Higgs self-coupling once we
take into account all operators and all uncertainties, so it hardly
serves as a motivation to measure a SM-like Higgs
self-coupling.\medskip

In this paper we take a slightly different approach. First, we assume
that the new physics responsible for the strongly first-order
electroweak phase transition only appears in the Higgs sector. In the
EFT framework we would consider, for example, the operator $(\phi^\dag
\phi)^3/\Lambda^2$~\cite{christophe_geraldine,eft2,Noble}.  While this approach
systematically includes higher-dimensional operators in a
power-counting expansion, it is not at all guaranteed that such an
expansion is appropriate for the underlying new physics. Furthermore,
a description of first-order phase transitions requires one to extract
global information about the effective potential. Again, a simple
polynomial expansion around a vanishing Higgs field 
might not be sufficient to resolve the fluctuation-driven competition
between different minima of the effective potential that induce a
first-order phase transition.

A simple global approximation to the effective potential is provided
by mean-field theory, which works remarkably well for Standard Model
parameters~\cite{mean-field,Gies:2013fua,Borchardt:2016xju,Sondenheimer:2017jin}
because of the dominance of the top quark. Depending, however, on the
strength of the bosonic and order-parameter fluctuations in the new
physics model, mean-field approaches may become unreliable.  We demonstrate this explicitly
using a simple example case in this paper.  This
situation calls for non-perturbative methods. Recently, lattice
simulations have been used to study the possibility of first-order
phase transition in the presence of the operator 
$(\phi^\dag\phi)^3/\Lambda^2$, both in a Higgs-Yukawa
model~\cite{Akerlund:2015fya} and in a gauged-Higgs
system~\cite{Akerlund:2015gfy}.

Here we use the functional renormalization group
(FRG)~\cite{christof_eq} as a non-perturbative tool, for reviews see,
\eg, \cite{rg_reviews}. It is able to provide global information
about the Higgs potential, bridge a wide range of scales, include
fluctuations of bosonic and fermionic matter fields as well as gauge
bosons and deal with extended classes of Higgs potentials. The two
questions which will guide us are:
\begin{enumerate}
\setlength{\itemsep}{0mm}
\item Do extended Higgs potentials help with electroweak baryogenesis?
\item Can they be systematically tested by measuring the Higgs self-coupling?
\end{enumerate}
We study the influence of operators or functions of operators in the
Higgs sector on the electroweak phase transition using several
representative examples. We determine the consequences for
the Higgs self-coupling for suitable extended Higgs potentials
supporting electroweak baryogenesis and being compatible with the
standard-model mass spectrum.\medskip

The global properties of the Higgs potential are also intimately
related to the questions of vacuum stability and Higgs mass
bounds~\cite{Krive:1976sg,Lindner:1985uk,Buttazzo:2013uya}. In fact,
higher-dimensional operators can also increase the stability regime of the
vacuum~\cite{Branchina:2005tu,Gies:2013fua,eichhorn_scherer,our_paper,Akerlund:2015fya,stable_frg,Jakovac:2015kka}.
The example Higgs potentials studied in this paper suggest new-physics
scales well below a possible instability scale of 
$10^{10\cdots12}$~GeV of the Standard Model. While vacuum instability
is therefore not an issue for our study, extended potentials generally
do have the potential to both support electroweak baryogenesis and
stabilize the Higgs vacuum. A measurement of the Higgs self-coupling
can therefore be indicative for both aspects.

\subsection{Electroweak phase transition}
\label{sec:phase}

The asymmetry between the matter and anti-matter contents in the
Universe is one of the great mysteries in cosmology and particle
physics. Experimentally, the effective absence of anti-matter in the
Universe has been proven in many different
ways~\cite{antimatter_meas}. A quantitative measurement is given by
the baryon-to-photon ratio $n_B/n_\gamma \approx 6 \cdot 10^{-10}$,
which is many orders of magnitude larger than what we would expect
from the thermal history in the presence of anti-matter. It can be
explained by a small initial asymmetry in the number of baryons and
anti-baryons which leads to a finite density of baryons after
essentially all anti-baryons have annihilated away.

Theoretically, the mechanisms behind the baryon asymmetry are well
understood. Most notably, it can be shown that the presence of an
asymmetry is equivalent to the three Sakharov conditions for our
fundamental theory~\cite{ew_phase}: baryon number violation, C as well
as CP violation, and departure from thermal equilibrium. The first two
conditions can be probed by precision measurements of the Lagrangian
of the Standard Model and its extensions.  The third condition can in
principle be achieved at the time of the electroweak phase transition,
where it then requires a strong first-order phase transition. The
nature of the electroweak phase transition can be read off from the scalar
potential in or beyond the Standard Model.

The strength of the phase transition which occurs at the critical
temperature $T_c$ is measured by the ratio $\phi_c/T_c$, where
$\phi_c=\langle \phi\rangle_{T_c}$ is the expectation value of the
Higgs at the critical temperature. The critical temperature describes
the transition where for small temperatures $T <T_c$ the potential
exhibits a single, non-trivial minimum for some value of the scalar
field $\phi$.  The field value at the minimum is temperature
dependent, approaching $v=246$~GeV for $T \rightarrow 0$.  With
increasing temperature, a second minimum at zero field value and with
an unbroken electroweak symmetry appears in a first-order scenario.
At the critical temperature $T_c$, the two minima of the potential,
\ie the one at finite field value and the one at vanishing field value
are degenerate, and the system undergoes a phase transition from the
symmetry-broken regime with a finite Higgs expectation value to the
symmetric regime.

The field value at the minimum constitutes an order parameter.  For
$\phi_c \neq 0$ the transition is of first order, \ie the vacuum does
not evolve continuously through the phase transition. For electroweak
baryogenesis, the transition has to be a strong first-order one,
\begin{equation}
  \frac{\phi_c}{T_c} \gtrsim 1 \,,
\end{equation}
otherwise the baryon asymmetry is washed
out~\cite{Shaposhnikov:1991cu}.

\subsection{Higgs self-coupling measurement}
\label{sec:self}

\begin{figure}[b!]
  \includegraphics[width=0.65\textwidth]{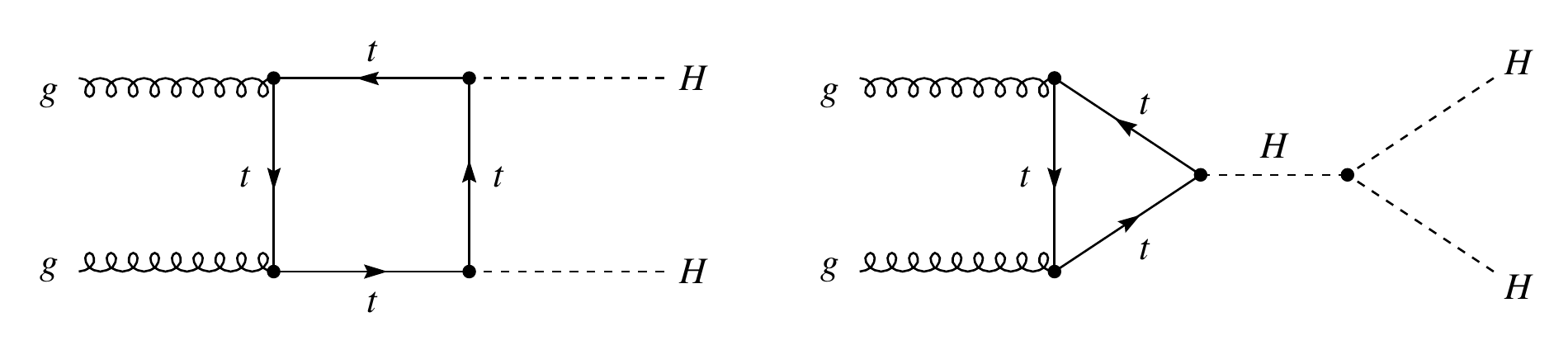}
\caption{Feynman diagrams contributing to Higgs pair production at the
  LHC. Figure from Ref.~\cite{uli1}.}
\label{fig:feynman}
\end{figure}

At energy scales relevant for the LHC, the self-interaction of the
Higgs boson is described by the infrared (IR) Higgs potential in the broken
phase. In the renormalizable Standard Model, and ignoring Goldstone
modes, it reads at tree level
\begin{align}
 V = \frac{\mu^2}{2} \, (v+H)^2 + \frac{\lambda_4}{4} \, (v+H)^4 \; ,
\label{eq:pot_ir}
\end{align}
where $H$ is the physical Higgs field.
The two parameters describing the SM-Higgs potential in the IR,
$\mu$ and $\lambda_4$, can be traded for the vacuum expectations value $v$ and
the Higgs mass $m_H$~\cite{lecture}
\begin{align}
 v = \sqrt{ \frac{\mu^2}{2 \lambda_4} } = 246~\gev\,, \qqqquad 
 m_H = \sqrt{2 \lambda_4} \, v = 125~\gev \; .
\label{eq:ir_values}
\end{align}
The interaction between three and four physical Higgs bosons in the
Standard Model is then given by
\begin{align}
\lambda_{H^3,0} = \frac{3 m_H^2}{v}\,, \qqqquad 
\lambda_{H^4,0} = \frac{3 m_H^2}{v^2} \; .
\label{eq:ir_selfs}
\end{align}
In the limit of heavy top quarks, $2 m_t > m_H$, an effective
Higgs--gluon Lagrangian~\cite{low_energy}
\begin{align}
\lag_{ggH} 
=\frac{\alpha_s}{12\pi} \; G^{\mu\nu}G_{\mu\nu} \;
 \log \left(1+\frac{H}{v} \right) 
= \frac{\alpha_s}{12 \pi} \; G^{\mu\nu}G_{\mu\nu} \; \frac{1}{v} \;
\left( H - \frac{H^2}{2v} + \ldots \right)\,,
\label{eq:higgs_eff}
\end{align}
with the gluon field strength tensor $G_{\mu\nu}$ and the strong coupling $\alpha_s$,
can be used to describe many relevant LHC observables.\medskip

When we include new physics contributions in the Higgs potential, the
relations in Eq.\eqref{eq:ir_values} change. It is instructive to
follow the simple example of the modified Higgs
potential~\cite{lecture}
\begin{align}
 V = \frac{\mu^2}{2} \, (v+H)^2 
   + \frac{\lambda_4}{4} \, (v+H)^4 
   + \frac{\lambda_6}{\Lambda^2} \, (v+H)^6 \; .
\label{eq:mod-Higgs-pot}
\end{align}
The modified relations between the observables become
\begin{align}
 m_H &= \sqrt{2 \lambda_4} \, v \left( 1 +12 \frac{\lambda_6 v^2}{ 
\lambda_4 \Lambda^2} \right) \,,\notag \\[1ex]
 \lambda_{H^3} &= \frac{3 m_H^2}{v} \left( 1 + \frac{16 \lambda_6 v^4}{ 
m_H^2 \Lambda^2} \right) 
              \equiv  \lambda_{H^3,0} \left( 1 + \frac{16 
\lambda_6 v^4}{ m_H^2 \Lambda^2} \right)\,, \notag \\[1ex]
 \lambda_{H^4} &= \frac{3 m_H^2}{v^2} \left( 1 +
 \frac{96 \lambda_6 v^4}{m_H^2 \Lambda^2} \right) 
               \equiv  \lambda_{H^4,0} \left( 1 +
 \frac{96 \lambda_6 v^4}{m_H^2 \Lambda^2} \right) \; .
\label{eq:ir_shifts}
\end{align}
Because $m_H$ and $v$ have to keep their measured values, we need to
adjust $\lambda_4$ to compensate for the effect of $\lambda_6$ on the
Higgs mass. This shift has to be accounted for in the expressions for
the Higgs self-couplings as a function of $m_H$ and $v$. The reference
couplings $\lambda_{H^n,0}$ keep their Standard Model values in terms
of the unchanged parameters $m_H$ and $v$, but the physical Higgs
couplings $\lambda_{H^n}$ change.\medskip

The standard channel to measure $\lambda_{H^3}$ at the LHC is Higgs
pair production in gluon fusion, as illustrated in
Fig.~\ref{fig:feynman},~\cite{orig,spirix,uli1,uli2,uli3,review_hh}.
Its production rate is known including NLO~\cite{nlo} and
NNLO~\cite{nnlo}.  One of the problems with such a measurement is that
the link between the total di-Higgs production rate and the Higgs
self-coupling requires us to know the top Yukawa coupling. An
appropriate framework is the global Higgs
analysis~\cite{legacy,hh_d6}, which is expected to give at best a 10\%
measurement of the top Yukawa coupling.  A model-independent precision
measurement of the top Yukawa coupling at the percent level will only
be possible at a 100~TeV collider~\cite{nimatron_yt}.

The experimental situation improves once we include kinematic
information in the di-Higgs production process.  Two kinematic regimes
are well known to carry information on the Higgs self-coupling, both
exploiting the (largely) destructive interference between the two
graphs shown in Fig.~\ref{fig:feynman}. While the continuum
contribution dominates over most of the phase space, the two diagrams
become comparable close to threshold~\cite{spirix,uli1}. The
low-energy theory of Eq.\eqref{eq:higgs_eff} gives us for the combined
di-Higgs amplitude
\begin{align}
\mathcal{A} \propto 
\frac{\alpha_s}{12 \pi v} \; 
\left( \frac{\lambda_{H^3}}{s-m_H^2} - \frac{1}{v} \right) 
\stackrel{\lambda_{H^3} = \lambda_{H^3,0}}{\longrightarrow}
\frac{\alpha_s}{12 \pi v^2} \; 
\left( \frac{3m_H^2}{3m_H^2} -1 \right) = 0 
\qquad \text{for} \quad 
m_{HH} \to 2 m_H \; ,
\label{eq:higgs_pair}
\end{align}
where $m_{HH}$ is the invariant di-Higgs mass.  An exact cancellation
occurs in the Standard Model.  Whereas the heavy-top approximation is
known for giving completely wrong kinematic distributions for Higgs
pair production~\cite{uli1}, it does correctly predict this threshold
behavior. Note, that the momenta of the outgoing particles in such processes are 
typically small compared to the Higgs mass and the low-energy regime of 
the theory is probed. In the analysis in Sec.~\ref{sec:main}, we thus read off 
the Higgs self-couplings from the low-energy effective potential.

The second relevant kinematic regime is boosted Higgs pair
production~\cite{boosted}, because of top threshold contributions to
the triangle diagram around $m_{HH} = 2 m_t$. In terms of the
transverse momentum this happens around $p_{T,H} \approx 100$~GeV,
where the combined amplitude develops a minimum for large Higgs
self-couplings.\medskip

At the LHC, we define di-Higgs signatures simply based on Higgs decay
combinations. The most promising channel is the $b\bar{b}\gamma\gamma$
final state~\cite{uli3,madmax,vernon,atlas}, where we can easily
reconstruct one of the two Higgs bosons and measure the continuum
background in the side bands.  We can also use the $b\bar{b} \tau
\tau$ final state~\cite{uli2,boosted}, assuming very efficient
tau-tagging.  The combination $b\bar{b}WW$~\cite{bbww} requires an
efficient suppression of the $t\bar{t}$ background, while the
$4b$~\cite{uli2,bbbb} and $4W$~\cite{uli1,wwww} signatures are
unlikely to work for SM-like Higgs bosons. Finally, the $b\bar{b} \mu
\mu$ is in many ways similar for the $b\bar{b} \gamma \gamma$
channel~\cite{uli3}, but with a much lower rate in the Standard Model.

To get an idea of what to expect, we quote the optimal reach of the
high-luminosity LHC run with $3~\iab$, based on the Neyman-Pearson
theorem applied to the $b\bar{b} \gamma \gamma$ channel for
self-couplings relatively close to the Standard Model~\cite{madmax},
\begin{alignat}{9}
\frac{\lambda_{H^3}}{\lambda_{H^3,0}} &= \phantom{-} 0.4~...~1.7 \qquad \text{at 68\% CL,}
\label{eq:lhc}
\end{alignat}
so any value for $\lambda_{H^3}/\lambda_{H^3,0}$ outside the range given  above will not be 
compatible with the vanishing di-Higgs amplitude in Eq.\eqref{eq:higgs_pair}.
This reach will be improved when we combine several Higgs decay
channels, but will also suffer from systematic uncertainties. In
addition, it assumes a perfect knowledge of the top Yukawa
coupling. This implies that models which predict a change in the
Higgs self-coupling by less than 50\% will not be testable at the LHC.

\section{Modified Higgs potentials}
\label{sec:model}

Similar to the EFT approach we assume that beyond an ultraviolet (UV)
scale or cutoff scale $\Lambda$ new physics exists and modifies the
form of the Higgs potential.  As the additional degrees of freedom are
heavy, their effects below $\Lambda$ can be parametrized by additional
terms in the Higgs potential, without modifying the propagating
degrees of freedom.  The details of the new physics are encoded in the
initial condition for the RG flow of the Standard Model at
$k =\Lambda$.  Exploring different higher-order terms thus provides
access to large classes of high-scale physics scenarios, for which we
do not have to investigate the detailed matching of the additional
terms in the Higgs potential and the underlying high-scale degrees of
freedom at $k = \Lambda$.

Our system features three relevant energy scales. First, the RG scale
$k$ ranges between $k=0$, where all quantum fluctuations are taken
into account, and $k=\Lambda$, where we initialize the flow. Second,
the temperature $T$ defines the external physics scale with which we
probe our system. Third, the field value $\phi$ defines an additional,
internal energy scale of our system. As is usual in EFT analyses, it is
important to clearly disentangle these three scales, even though $\phi$
and $T$ can in principle act similarly to the RG scale $k$ in that
they suppress IR quantum fluctuations~\cite{our_paper}. We employ a method
that can straightforwardly account for the RG flow in the presence of
these different scales, namely the functional renormalization group.
In this setting, quantum fluctuations in the presence of further
internal and external scales are taken into account by a functional
differential equation that is structurally one-loop, without being
restricted to a weak-coupling regime. This provides access to classes
of non-perturbative microscopic models with a manageable computational
effort. Most importantly, the functional RG approach enables us to
keep track of the separate dependence of the potential on the RG scale
$k$, the temperature and the field value even in cases with
non-perturbative UV potentials, where, \eg a mean-field approach
breaks down.\medskip

For our study, we concentrate on that part of the Standard Model which
is relevant for the RG flow of the Higgs potential using the framework
developed in \cite{our_paper}. Here, we follow  that framework by 
implementing the effects of weak gauge bosons through a fiducial coupling, 
and upgrade our treatment by including a thermal mass generated by the
corresponding fluctuations as their leading contribution instead of 
implementing a fully-fledged dynamical treatment of that sector, 
see App.~\ref{app:der-flows} for details.
Similarly, would-be Goldstone modes do not
need to be considered explicitly, such that it suffices to concentrate
on a real scalar field $\phi$, which after electroweak symmetry
breaking can be described in terms of the physical Higgs field $H$ as
$\phi = H + v$.  At the UV scale $k=\Lambda$, the Higgs-potential is
parametrized as
\begin{align} 
 V_{k=\Lambda} = \frac{\mu^2}{2} \, \phi^2 + 
\frac{\lambda_4}{4} \, \phi^4  + \Delta V \,,
\label{eq:poly-start-pot}
\end{align}
where $\Delta V$ contains the contribution of some higher dimensional
operator. In principle, higher-order modifications of the Yukawa
sector could also be included,
cf.~\cite{Pawlowski:2014zaa,deVries:2017ncy,Gies:2017zwf}. We investigate three
classes of modifications to the SM-Higgs potential:
\begin{enumerate}
\setlength{\itemsep}{0mm}
\item additional $\phi^6$ or $\phi^8$ terms, which cover the
  leading-order terms in an effective-field theory approach and have
  been extensively studied in the
  literature~\cite{eft1,christophe_geraldine,eft2,Noble};
\item a logarithmic dependence on the Higgs-field, inspired by
  Coleman-Weinberg potentials. It does not allow for a Taylor expansion
  around $\phi=0$. Logarithmic modifications are naturally
  generated by functional determinants, \ie by integrating out
  heavy scalars or fermions.
\item a simple example of non-perturbative contributions of the form
  $\exp(-1/\phi^2)$, \ie an exponential dependence on the inverse
  field, consequently not admitting a Taylor expansion in the field
  around $\phi=0$. This is inspired by semiclassical contributions to
  the path integral with $\phi$ reminiscent to a moduli parameter of
  an underlying model.
\end{enumerate}
We denote these modifications of the potential by
\begin{alignat}{9} 
 \Delta V_6  &= \lambda_6 \, \frac{\phi^6}{\Lambda^2}\,,  \qqqquad &
 \Delta V_8  &= \lambda_6 \, \frac{\phi^6}{\Lambda^2}
             +  \lambda_8 \, \frac{\phi^8}{\Lambda^4} \,, \notag\\[2ex]
 \Delta V_{\ln,2}  &= -\lambda_{\ln,2} \, \frac{\phi^2\Lambda^2}{100} \; 
\ln \frac{\phi^2}{2\Lambda^2}\,, \qqqquad &
 \Delta V_{\ln,4}  &= \lambda_{\ln,4} \, \frac{\phi^4}{10} \; \ln \frac{\phi^2}{2\Lambda^2} 
\,, \notag\\[2ex]
 \Delta V_{\exp,4}  &= \lambda_{\exp,4} \phi^4  \exp\left(-
\frac{2\Lambda^2}{\phi^2}\right)\,, \qqqquad &
 \Delta V_{\exp,6}  &= \lambda_{\exp,6} \frac{\phi^6}{\Lambda^2}  
\exp\left(-\frac{2\Lambda^2}{\phi^2} \right)\,.
\label{eq:pots}
\end{alignat}
In all these potentials $\Lambda$ describes a new physics scale, which
absorbs the mass dimension of the Higgs field.  The case of
$\phi^6/\Lambda^2$ has been explored in the
literature~\cite{eft2,Noble,christophe_geraldine} and serves as a test
of our method, as discussed in the Appendix.  Neither the logarithmic
nor the exponential potentials can be expanded around $\phi=0$, so
they cannot be treated in an EFT framework. Similar bare potentials 
have been suggested in \cite{Sondenheimer:2017jin} in the context of 
Higgs mass bounds and vacuum stability. Instead, all potentials
that can be expanded around $\phi=0$ can be approximated by the
power-ordered, first kind of potentials. As expected by canonical
power counting, terms of higher order in $\phi$ can only play a role
for very low values of $\Lambda/v$, unless their prefactors are
non-perturbatively large. From a more general viewpoint, the set of
power law, logarithmic and exponential potential functions does not
only reflect the physics structures arising from local vertex
expansions, one-loop determinants or semiclassical approximations. It
also includes the set of functions to be expected on mathematical
grounds if the effective potential permits a potentially
resurgent transseries expansion \cite{Dunne:2012ae}.\medskip

\begin{figure}[t]
 \centering
 \includegraphics[width=0.5\textwidth]{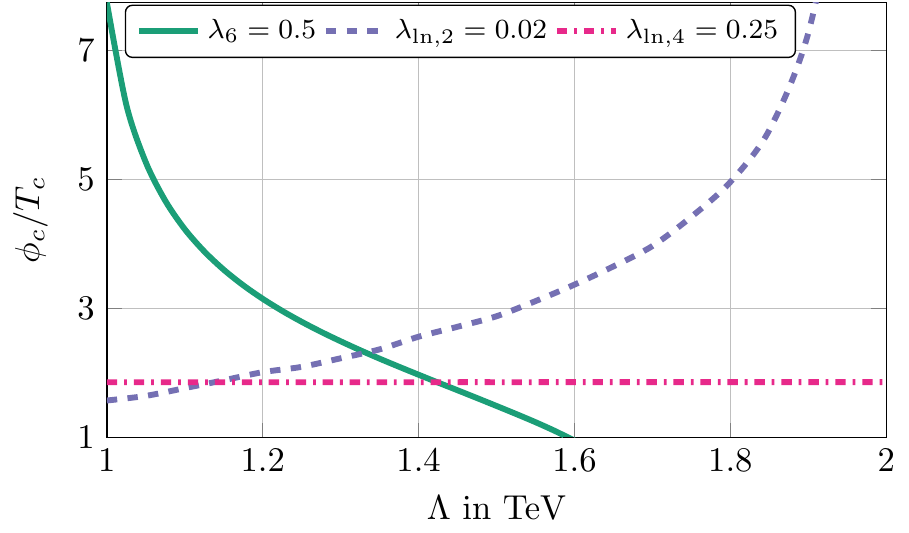}
 \caption{Mean-field results for $\phi_c/T_c$ as a function of the cutoff
   for different modifications of the Higgs potential. Second-order and 
   weak first-order phase transitions are excluded from the plot.
   The results of the
   $\phi^6$ modification are reasonable, while the results for the 
   $\phi^2\ln\phi^2$ and the $\phi^4\ln\phi^2$ modifications are clearly
   unphysical, see explanation in the text. More elaborate methods than mean-field are needed.
   }
 \label{fig:mean-field}
\end{figure}

To investigate the different classes of modifications, a variety of
tools appears to be at our disposal, a priori ranging from mean-field
techniques to non-perturbative lattice tools and functional methods. It
turns out that the former are only applicable to a restricted class of
potentials, not allowing us to adequately explore the full range of
possible UV potentials corresponding to diverse underlying microscopic
models. This is displayed in Fig.~\ref{fig:mean-field} where the $\phi^6$-
modification of the Higgs potentials shows the expected physical behavior
as the strength of the first-order phase transition is decreasing with 
an increasing cutoff. The logarithmic modifications on the other hand
show a rather unphysical behavior as the strength of the first-order
phase transition remains constant or even increases with the UV scale.
This indicates that scalar order-parameter fluctuations are important, 
which are ignored in simple mean-field theory.
Therefore we make use of powerful functional techniques, which treat bosonic
and fermionic fluctuations on the same footing.

When allowing for modifications of the Higgs potential, we need to
ensure that at $T=0$ the IR-values for $\mu$, $\lambda_4$,
and the top-Yukawa-coupling $y_t$ are such that the 
measured observables do not change. We adjust the corresponding masses to
\begin{align}
 v = 246 \,\gev\,, \qqqquad 
 m_H = 125 \,\gev\,, \qqqquad 
 m_t = 173 \,\gev \,.
\label{eq:ir_data}
\end{align}
Within our numerical analysis, we require $v$ and $m_t$ to be
reproduced to an accuracy of $\pm$0.5~GeV. The Higgs mass is adjusted
within a somewhat larger numerical band of $\pm 1.5$~GeV. Since it is
related to the second derivative (curvature) of the
potential at the minimum, a higher precision is numerically more
expensive, see App.~\ref{app:grid-code} for details.
Moreover, it is expected that the curvature mass used here shows
small deviations from the Higgs pole mass $m_H$, see
\cite{Helmboldt:2014iya}, and the above band also contains an
estimate of this systematic error.  In the symmetry broken regime,
the potential given in Eq.\eqref{eq:poly-start-pot} can be expanded in
powers of $(\phi^2 - v^2)$.  In the decoupling region in the deep IR,
we use the parametrization
\begin{align}
  V_{k\ll v} &= \frac{\lambda_{4,\,\rm IR}}{4} (\phi^2 - v^2)^2 
               +  \frac{\lambda_{6,\,\rm IR}}{8 v^2} (\phi^2 - v^2)^3
               +  \frac{\lambda_{8,\,\rm IR}}{16v^4} (\phi^2 - v^2)^4 + \cdots\notag \\
             &=  \lambda_{4,\,\rm IR} v^2 H^2 
               + \left( \lambda_{4,\,\rm IR} + \lambda_{6,\,\rm IR} \right) v H^3 
               + \frac{1}{4} \left( \lambda_{4,\,\rm IR}  + 6 \lambda_{6,\,\rm IR}  +
 4 \lambda_{8,\,\rm IR}   \right) H^4 
               + \cdots\,.
\end{align}
Note that this is the full effective potential in the IR, differing
from the tree-level potential in Eq.\eqref{eq:mod-Higgs-pot}. In
particular, higher-order terms, encoded in $\lambda_{6,\, \rm IR }$
are generated by quantum fluctuations even if the tree-level potential
is quartic.  At tree level, the Higgs potential is described by two
parameters, \ie
$\lambda_{6,\,\rm IR} = \lambda_{8,\,\rm IR} = \ldots =0$.  If we
allow higher-order terms, all measurable parameters are affected, in
close analogy to Eq.\eqref{eq:ir_shifts}.  As described in
Sec.~\ref{sec:self} the vacuum expectation value $v$ and the Higgs
mass $m_H^2/(2v^2) \equiv \lambda_4$ are known very precisely from
collider measurements and thus we have to keep them fixed.  The
physical Higgs self-couplings change from the values given in
Eq.\eqref{eq:ir_selfs} to the more general form
\begin{align}
 \lambda_{H^3} = \frac{\delta^3}{\delta H^3} V_{k=0} 
	       =  6 v(\lambda_{4,\,\rm IR} + \lambda_{6,\,\rm IR})\,, \qquad 
 \lambda_{H^4} = \frac{\delta^4}{\delta H^4} V_{k=0} 
	       = 6 (\lambda_{4,\,\rm IR}  + 6 \lambda_{6,\,\rm IR}  + 4 \lambda_{8,\,\rm IR})  \; .
\end{align}
The first terms are precisely the couplings
$\lambda_{H^3,0}=6 v\lambda_{4,\,\rm IR}$ and
$\lambda_{H^4,0}=6 \lambda_{4,\,\rm IR}$ familiar from the tree-level
structure.  With the present setup we can compute the Higgs
self-couplings in the pure Standard Model including higher-order terms
generated by quantum fluctuations by initializing the flow at some
high cutoff scale without any modifications of the Higgs potential. As
long as the cutoff is not too close to the electroweak scale the
results will be largely independent of the cutoff choice. For our
level of numerical precision, a cutoff $\Lambda=2$~TeV is sufficient.
The Higgs self-couplings are given by
\begin{align}
 \frac{\lambda_{H^3}}{\lambda_{H^3,0}} \approx 0.92\,, \qqqquad 
 \frac{\lambda_{H^4}}{\lambda_{H^4,0}} \approx 0.68 \;.
 \label{eq:Higgs-self-SM}
\end{align}
These values are equivalent to computations of the Higgs potential
with Coleman-Weinberg corrections.  We then go beyond the pure
Standard Model by adjusting a combination of the coefficients
$\lambda_j$ and the new physics scale $\Lambda$ in
Eq.\eqref{eq:pots}. These can now be used to adjust $\phi_c/T_c$ such
that we obtain a strong first-order phase transition.

\section{Phase transition}
\label{sec:main}

For the modified Higgs potentials defined in Eq.\eqref{eq:pots} we
need to explore which values of the UV scale $\Lambda$ and the
coefficients $\lambda_j$ lead to a sufficiently strong first-order
transition. Simultaneously, we monitor whether this leads to a
measurable modification of the Higgs self-couplings in the~IR.

\subsection{First-order phase transition}
\label{sec:pt}

\begin{figure}[t]
 \centering
 \includegraphics[width=0.6\textwidth]{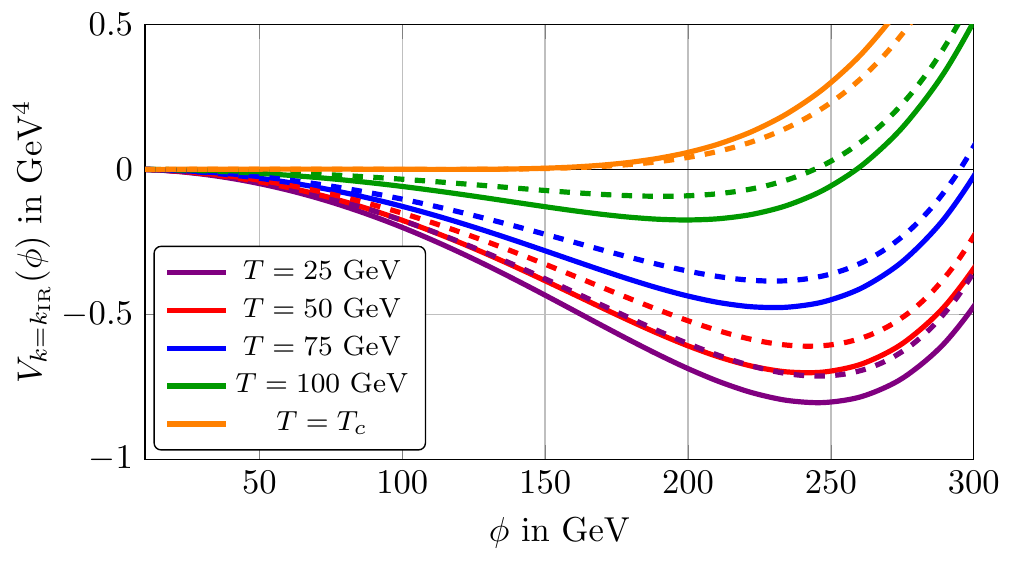}
 \caption{Temperature evolution of the potentials of the type
   $\phi^4 \ln\phi^2$ (solid) and $\phi^4 \exp(-1/\phi^2)$ (dashed)
   for fixed $\phi_c/T_c\approx1$.  We plot the temperatures $T=25$
   GeV (violet), $T=50$ GeV (red), $T=75$ GeV (blue), $T=100$ GeV
   (green) and $T=T_c$ (orange). Note that
   $T_c^{\ln,4} = 116.4\;\mathrm{GeV} > T_c^{\exp,4} = 110.5$ GeV and
   thus one curve overtakes the other.  A magnification of the curves
   at $T=T_c$ is displayed in Fig.~\ref{fig:IR-plots} }
 \label{fig:T-comp}
\end{figure}

\begin{figure}[t]
 \includegraphics[width=\textwidth]{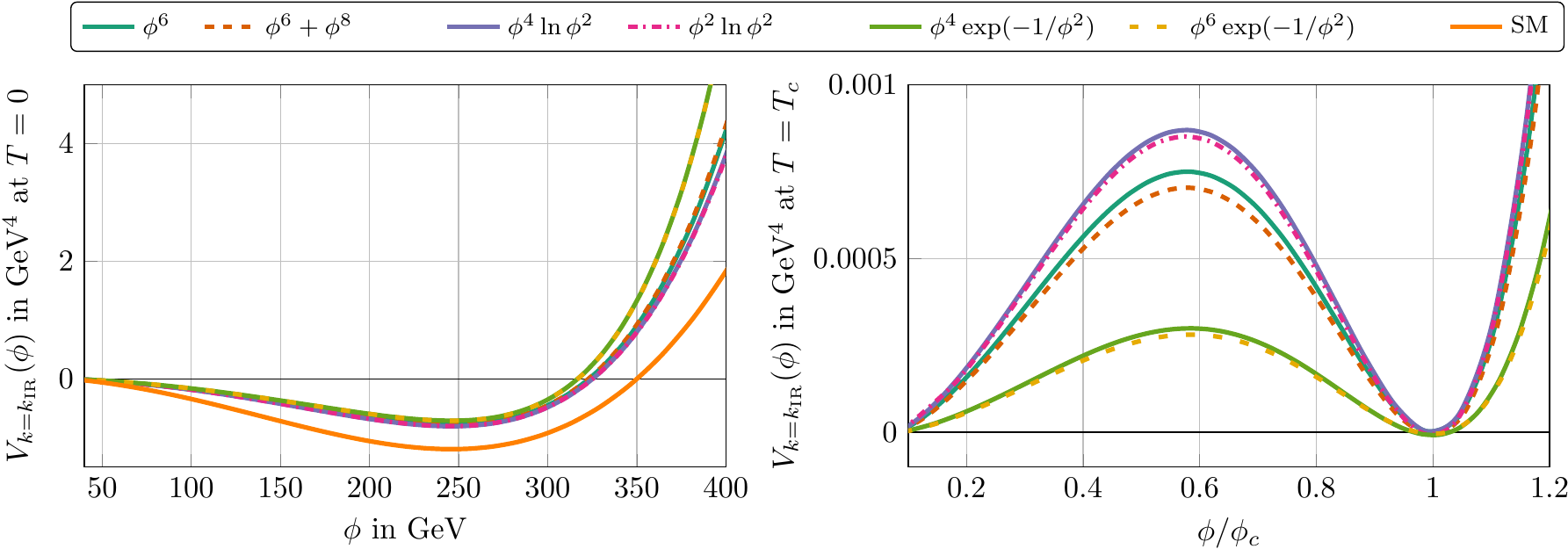}
 \caption{Effective potentials at $T=0$ (left) and $T=T_c$ (right). We
   show all modified Higgs potentials from Eq.\eqref{eq:pots} with
   $\Lambda=2$~TeV. The values of the coefficients at the UV scale
   $\Lambda$ are fixed by the requirement $\phi_c \approx T_c$,
   leading to $\lambda_6=1.2$, $\lambda_6=1$ with $\lambda_8=1.4$,
   $\lambda_{\ln,4}=0.89$, $\lambda_{\ln,2}=0.27$,
   $\lambda_{\exp,4}=23.3$, and $\lambda_{\exp,6}=27.5$.}
 \label{fig:IR-plots}
\end{figure}

In Fig.~\ref{fig:T-comp} we show the evolution of two example
potentials from Eq.\eqref{eq:pots} from zero temperature to $T_c$,
where the latter is defined as the temperature at which the two
competing minima become degenerate. The latter is not distinctly
apparent in Fig.~\ref{fig:T-comp}, but becomes visible in the
magnification in the right panel of Fig.~\ref{fig:IR-plots}.  We also
require the second minimum to be at $\phi_c= T_c$, to guarantee a
sufficiently strong first-order phase transition. This way, the $\phi$
dependence of the two cases becomes comparable.  A key feature already
visible in this figure is that the potential with the deeper minimum
at small temperature turns into the steeper potential at $T_c$.  This
is achieved by a larger value of $T_c$ for the potential with the
deeper minimum.
Note that the potentials in Fig.~\ref{fig:T-comp} and \ref{fig:IR-plots}
are read off at the RG scale $k_\text{IR}$, which is an infrared scale where
the Higgs potential and all observables are frozen out. Below this
scale only convexity generating processes take place. The freeze out occurs once 
fluctuations of fields decouple from the RG flow because the RG scale $k$ crosses their 
mass-threshold. This decoupling is built into the FRG setup. We choose $k_\text{IR}$
to be smaller than the masses of the model, such that the exact choice of $k_\text{IR}$
does not matter.

In Fig.~\ref{fig:IR-plots} we illustrate the behavior of all our
modified Higgs potentials in the IR at vanishing temperature (left
panel) and at the critical temperature (right panel), respectively.
Note the different scales on the vertical axes. The UV scale $\Lambda$
and the respective coefficients $\lambda_j (\Lambda)$ are chosen such
that they result in a strong first-order phase transition,
$\phi_c/T_c=1$.  The different potentials at zero temperature are
similar to that of the Standard Model, as expected from the fact that
we fix the Higgs vacuum expectation value and mass to their observed
values. In particular, the minima all appear at $v = 246$~GeV, and the
second derivatives have to reproduce the measured Higgs mass.
Nevertheless, if we fix $V_{k=k_\text{IR}}(0) = 0$, an imprint of
modified UV physics remains visible.

In the left panel of Fig.~\ref{fig:IR-plots} we see that up to
$\phi \approx 300$~GeV, all modifications we consider lead to a very
similar form of the zero-temperature IR potential, if their
coefficients are fixed such that $\phi_c/T_c$ is the same for all our
potentials. At higher field values the different UV modifications lead
to distinct field-dependence of the potential. The sizable impact of
the modified microscopic action on the IR potential is due to the
finite UV scale $\Lambda = 2$~TeV. This is not sufficiently far above
the electroweak scale for the contributions $\Delta V$ to be washed
out by the RG flow. \medskip

At finite temperature, we see in the right panel of
Fig.~\ref{fig:IR-plots} that the potentials show significant
deviations and the six different modifications fall into three
distinct forms of the IR potential at $T_c$.  The Standard Model is
not displayed, since it exhibits a second-order phase transition with
$\phi_c=0$. The other potentials show different sizes of the bump
that separates the minima at $\phi=0$ and $\phi=\phi_c$. The
exponential modifications show the smallest bump, while logarithmic
modifications show the largest bump.  The third class is given by the
polynomial UV potentials, which fall in between the two other classes. \\ 

It is worth noting that the resulting IR modifications almost coincide
\emph{within} each class of UV potentials, \ie, the polynomial,
logarithmic, and exponential class. Although there are manifestly
different UV modifications within each class, like for instance
$\phi^4\exp(-1/\phi^2)$ vs $\phi^6\exp(-1/\phi^2)$, the resulting
IR behavior appears to be dominated by the exponential dependence, and
accordingly is nearly the same for the two cases -- as stressed
before, the two exponential cases differ from the two logarithmic
cases, which are within a separate class of their own.

Comparing the two panels we observe that zero-temperature potentials
with a steeper increase at larger field values turn into more shallow
potentials for finite temperature near the broken vacuum.  The latter
corresponds to a lower barrier between the two minima. The reason for
this link is that the phase transition occurs once positive thermal
corrections to the mass parameter are large enough to change the
extremum at $\phi=0$ from a maximum to a minimum, which then becomes
degenerate with the minimum at a finite field value.  For potentials
with a lower zero-temperature depth --- and correspondingly a more
substantial slope at large $\phi$ --- the corresponding critical
temperature $T_c$ is lower.  Therefore, the steepest increase towards
large $\phi$ in the left panel in Fig.~\ref{fig:IR-plots} corresponds
to the smallest bump in the right panel of
Fig.~\ref{fig:IR-plots}. Phrased differently: for potentials with a
flatter inner region, scalar fluctuations are quantitatively more
relevant. At the same time, the phase transition turns first order as
soon as the scalar fluctuations dominate over the fermionic ones.
This connection will become important when evaluating the prospects of
the different cases with regards to detectability at the LHC.

\subsection{Scale of new physics}
\label{sec:valid}

Given a particular microscopic model containing additional degrees of
freedom, the UV scale or cutoff $\Lambda$ is typically identified with
the mass scale of those additional fields, below which their
fluctuations are suppressed.  From an EFT point of view, one
correspondingly associates $\Lambda$ with the energy scale, above
which new physics can appear as on-shell excitations. In turn, below
$\Lambda$ the effect of new physics is only visible indirectly. Such
an indirect effect would be a deviation of the Higgs potential from
its form in the renormalizable Standard Model. A key aspect of this
kind of approach is that an EFT description by definition comes with a
region of validity, above which we will be sensitive to the actual UV
completion. Hence, before we use our modified Higgs potential to link
a strong first-order phase transition to the Higgs self-coupling we
need to study the validity range of our description. 

\begin{figure}[t]
 \includegraphics[width=\textwidth]{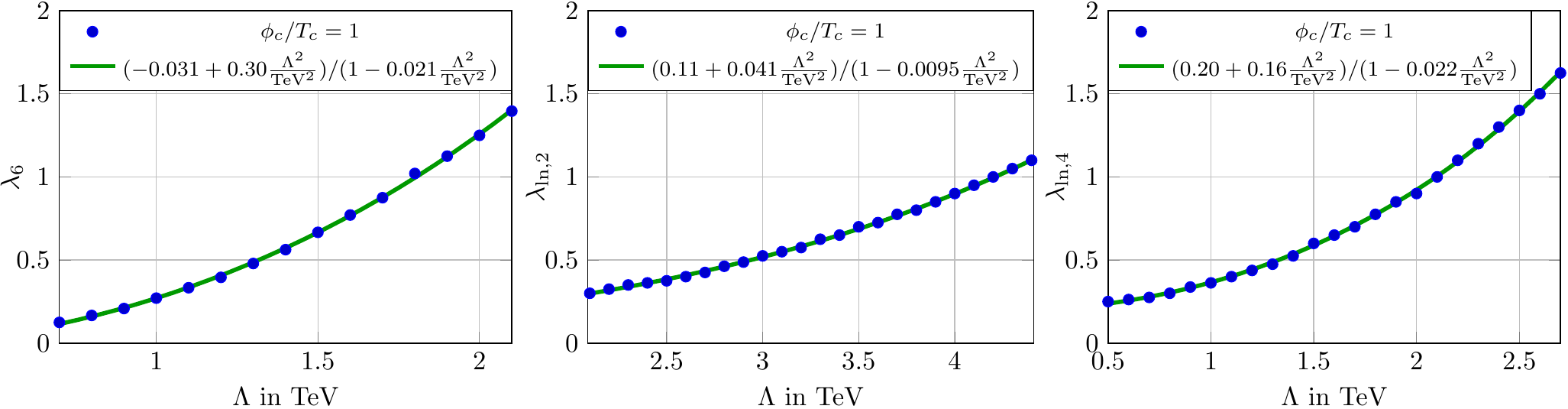}
 \caption{Coefficient $\lambda_j(\Lambda)$ of the dimension-6
   operator $\phi^6/\Lambda^2$ (left), the modification $\Lambda^2
   \phi^2 \ln \phi^2/\Lambda^2$ (center), and the modification $\phi^4
   \ln \phi^2/\Lambda^2$ (right) as a function of the cutoff,
   requiring $\phi_c/T_c = 1 \pm 0.05$.}
 \label{fig:pctc1}
\end{figure}

Following Eq.\eqref{eq:pots} we see that an indirect measurement using
an EFT-like approach is only sensitive to a combination of the scale
$\Lambda$ and the (Wilson) coefficients $\lambda_j$. In
Fig.~\ref{fig:pctc1} we show the correlation between
$\Lambda$ and the corresponding $\lambda_j$ evaluated at the UV scale
$\Lambda$ for a set of modified Higgs potentials, assuming a strong
first-order phase transition with $\phi_c/T_c = 1$.  We can interpret
these results as lines of constant IR physics: the running coefficient
$\lambda_j(\Lambda)$ then describes a family of effective models
defined at different scales $\Lambda$, all yielding the same IR observables.
Without new physics effects, $\Delta V = 0$, this corresponds
to fixing $v$, $m_H$ and $m_t$ in the IR and simply evolving them
toward the UV with their known RG equations. In our extended setup, the
additional coefficients measure the strength of the new physics
contribution, that we initialize at the UV scale $\Lambda$.  We then
use a corresponding parameter $\lambda_j$ to fix $\phi_c/T_c$ to a value of our choice.
Doing so for different UV scales $\Lambda$, the coefficient $\lambda_j$ becomes a function of
$\Lambda$.\medskip

Without running effects for the coefficients $\lambda_j$ the
correlation between the coefficient and the UV scale would be
simple. For instance, the dimension-6 Wilson coefficient would follow
a parabola, $\lambda_6 \propto \Lambda^2$. However, the condition on
$\phi_c/T_c$ for the strong first-order phase transition is defined at
energies around the Higgs VEV, while the shown values of $\lambda_j$
are defined in the UV.  The complete correlation is well-described by
a quadratic polynomial. In the case of $\lambda_6$, this reflects the
quadratic running due to the canonical dimension.  While the
normalization of $\Delta V$ can be adjusted at will and the absolute
values of the coefficients $\lambda_j$ do not carry any physical
significance, the growth of these coefficients towards the ultraviolet
suggests the possible onset of a strongly coupled regime.

To investigate the onset of this strongly coupled regime we fit the
correlation between $\lambda_j$ and $\Lambda$ to a broken rational
polynomial. A motivation for the particular choice of fit function in
Fig.~\ref{fig:pctc1} is given by an approach to a
power-like Landau-pole singularity. Indeed, this ansatz fits our numerical results well
for the given range of UV scales.  From the broken polynomial we can
estimate the critical scales, where the respective models might become
strongly coupled,
\begin{align}
 \Lambda_6^\text{crit} = 7.0\,\tev, \qqquad
 \Lambda_{\ln,2}^\text{crit} = 10 \,\tev, \qqquad
 \Lambda_{\ln,4}^\text{crit} = 6.8 \,\tev \,.
\end{align}
These critical scales should be viewed as conservative estimates of
the validity scale up to which our field-theory description using
purely Standard-Model degrees of freedom is applicable.  These
estimates are of the same order of magnitude as maximum values of
$\Lambda$ that lead to a first-order phase transition in studies based
on mean-field arguments, see \eg \cite{christophe_geraldine}.

\subsection{Baryogenesis vs Higgs self-coupling}
\label{sec:baryogenesis}

\begin{figure}[t]
 \includegraphics[width=\textwidth]{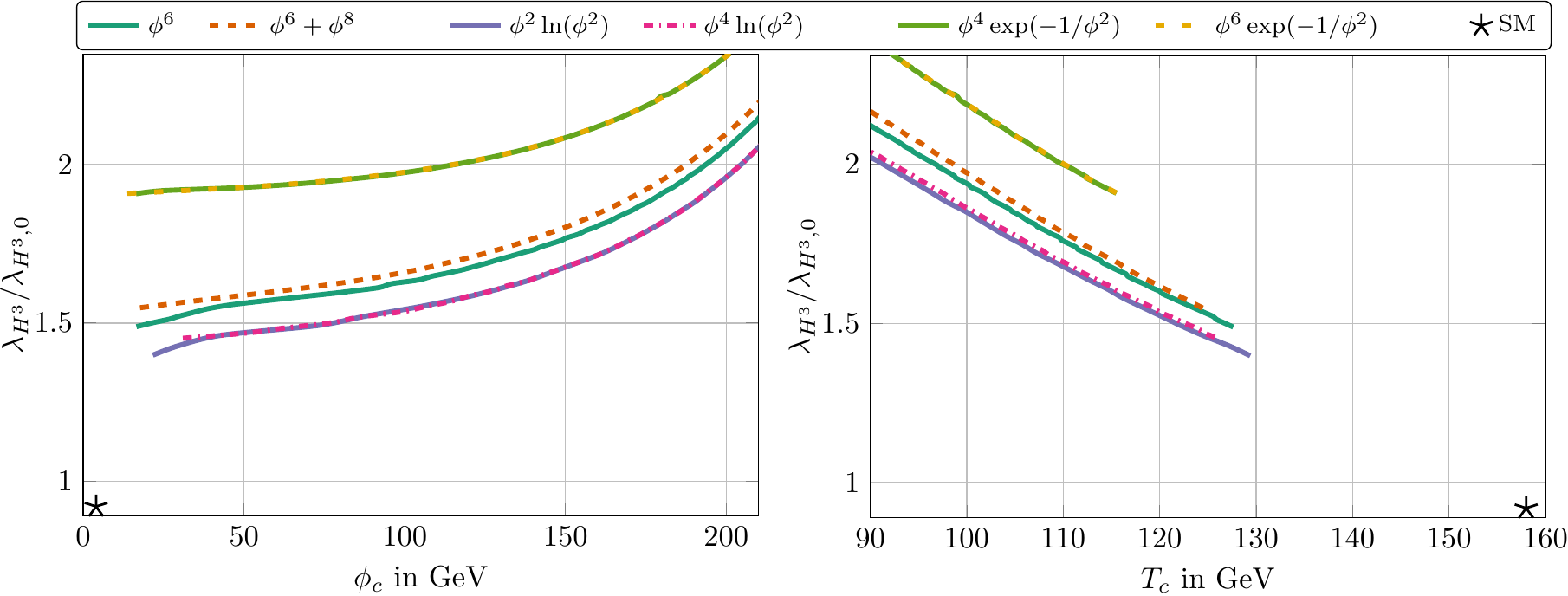}
 \caption{Modification of the self-coupling
   $\lambda_{H^3}/\lambda_{H^3,0}$ as a function of $\phi_c$ (left)
   and $1/T_c$ (right) for the UV potentials given in
   Eq.\eqref{eq:pots}.
   The asterisk in both plots represents the 
   Standard Model expectation, including Coleman-Weinberg corrections, 
   cf.~Eq.\eqref{eq:Higgs-self-SM}.}
 \label{fig:Higgs-self-sep}
\end{figure}

After showing how a modified Higgs potential can lead to a strong
first-order phase transition in Sec.~\ref{sec:pt} and confirming that
our approach is consistent in Sec.~\ref{sec:valid}, we can now explore
the link between the strong first-order phase transition and the
observable Higgs self-coupling. As laid out in the Introduction, the
crucial question is as to whether modifications of the Higgs potential that
lead to a sufficiently strong first-order phase transition for
electroweak baryogenesis can be tested through the Higgs self-coupling
measurement at the LHC.\medskip

Following the above discussion, the remaining question is how a value
$\phi_c/T_c \approx 1$ due to the potentials given in
Eq.\eqref{eq:pots} is reflected in shifted physical Higgs
self-couplings $\lambda_{H^3}$ and $\lambda_{H^4}$. All new physics
models are adjusted to reproduce the low-energy measurements in
Eq.\eqref{eq:ir_data}.  First, we can separate the two parameters
$1/T_c$ and $\phi_c$ and show their individual effects on the physical
Higgs self-couplings.  In Fig.~\ref{fig:Higgs-self-sep} we first see
that the two parameters contribute roughly similar amounts to an
increase in the Higgs self-couplings, if we push the model towards a
strong first-order phase transition. Second, we see that the
individual potentials in the general class of power-series,
logarithmic, and exponential potentials give essentially degenerate
results. Finally, the effect on the self-couplings is the weakest for
the logarithmic potential, slightly stronger for the power-law
modification, and the strongest for the exponential
modification.

\begin{figure}[!t]
 \includegraphics[width=\textwidth]{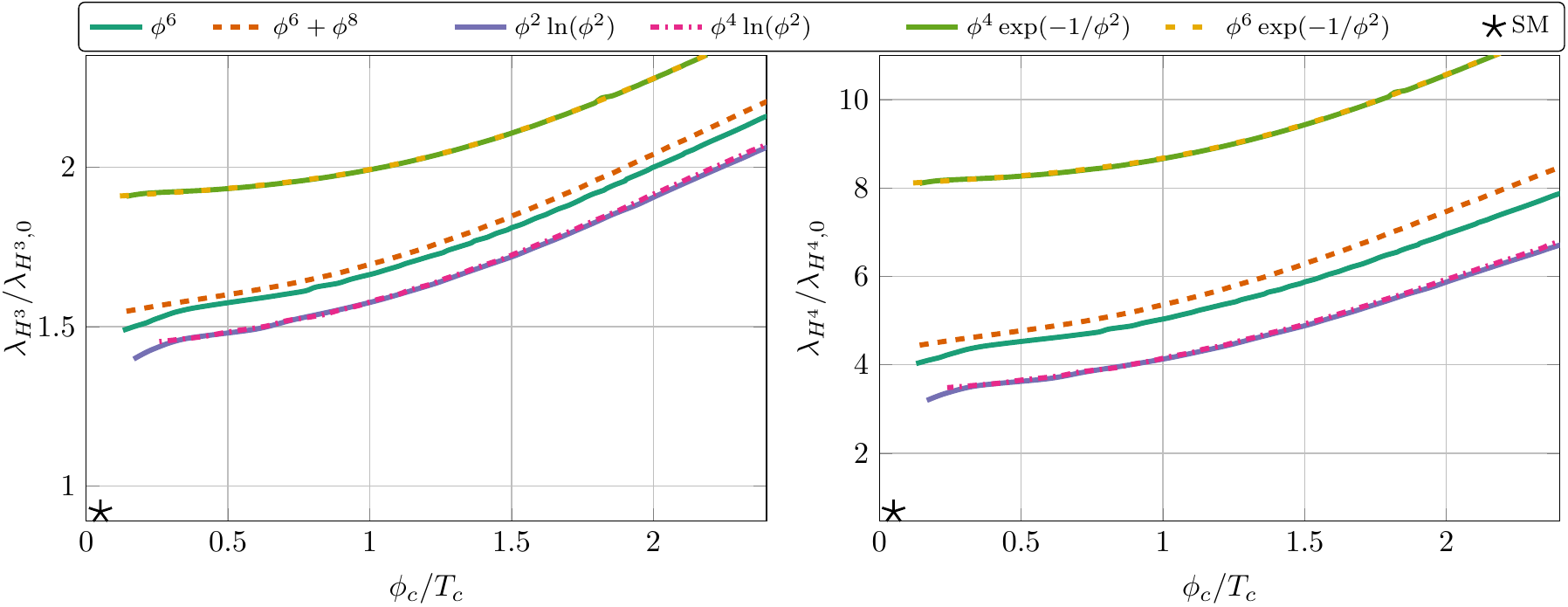}
 \caption{Modification of the self-couplings
   $\lambda_{H^3}/\lambda_{H^3,0}$ (left) and
   $\lambda_{H^4}/\lambda_{H^4,0}$ (right) as a function of
   $\phi_c/T_c$ for the UV potentials given in Eq.\eqref{eq:pots}.
   The asterisk in the lower left of both plots represents the 
   Standard Model expectation, including Coleman-Weinberg corrections, 
   cf.~Eq.\eqref{eq:Higgs-self-SM}.
   }
 \label{fig:Higgs-self}
\end{figure}

As already observed in Sec.~\ref{sec:pt}, a steeper zero-temperature
potential at large field values can be linked to a decrease in
$T_c$. On the other hand, a steeper increase at large field values
will be tied directly to larger values of the cubic and quartic Higgs
self-coupling. This dependence is confirmed by
Fig.~\ref{fig:Higgs-self-sep}, where potentials with smaller $T_c$
feature larger $\lambda_{H^3}$. This feature holds both within each
class of potentials where we can decrease $T_c$ by enhancing $\Delta
V$, and between different classes of potentials. This trend should be
generic in that additions $\Delta V$ leading to a strong first-order
transition at low $T_c$ will be easier to detect at the LHC.\medskip

Given that we do not see any striking effects from the
individual dependence on $1/T_c$ and $\phi_c$, we study the
dependence of the different Higgs potentials on the physically
relevant ratio $\phi_c/T_c$. In Fig.~\ref{fig:Higgs-self}, we show
the modifications of both Higgs self-couplings as a function of
$\phi_c/T_c$.  The free model parameter along the shown line is an
appropriate combination of new-physics scale $\Lambda$ and the
new-physics coefficient $\lambda_j$.  For $\phi_c/T_c \gtrsim 1$ we
find a strong first-order phase transition, suitable for electroweak
baryogenesis.
From the location of the Standard Model
point it is clear that there exists a range of modified self-couplings
where the electroweak phase transition remains second order. Only for
\begin{align}
\frac{\lambda_{H^3}}{\lambda_{H^3,0}} \gtrsim 1.5 
\qqquad \text{or} \qqquad 
\frac{\lambda_{H^4}}{\lambda_{H^4,0}} \gtrsim 4 \,,
\end{align}
we have a chance to generate a first-order phase transition.
This number should be compared to the LHC reach given in
Eq.\eqref{eq:lhc}. We conclude that the prospects of a detectable 
imprint appear to be good for all models that we have studied.
A strong first-order phase transition
corresponding to $\phi_c/T_c >1$ can in all scenarios be achieved by
further increasing the new physics contributions and thereby
increasing the Higgs self-couplings. In particular, we observe that
the non-perturbative modifications $\exp(-1/\phi^2)$ lead to a
significantly higher value of the Higgs self couplings at fixed
$\phi_c/T_c$ and are thus easier to detect. Given that for example
exponential potentials feature a minimum value of $\lambda_{H^3}$
significantly larger than the simple $\phi^6$ extension, the LHC
measurement might even allow first clues to
the nature of new physics,
even if the corresponding scale $\Lambda$ remains out of direct reach
at the LHC.\medskip

Because the curves in Fig.~\ref{fig:Higgs-self} connect an IR observable
with a UV property we can link the two regimes and make two
observations. First, we can start in the IR and fix $\lambda_{H^3}$
for different UV potentials. Here, we find that an increase in
$\phi_c/T_c$ or decrease in $T_c$ leads to a decrease in
$\lambda_{H^4}$ for constant $\lambda_{H^3}$. Alternatively, we can
fix $\phi_c/T_c$ for different UV potentials and find that a decrease
in $\lambda_{H^3}$ corresponds to a decrease also in $\lambda_{H^4}$
or an increase in $T_c$.\medskip

\begin{figure}[!t]
 \includegraphics[width=\textwidth]{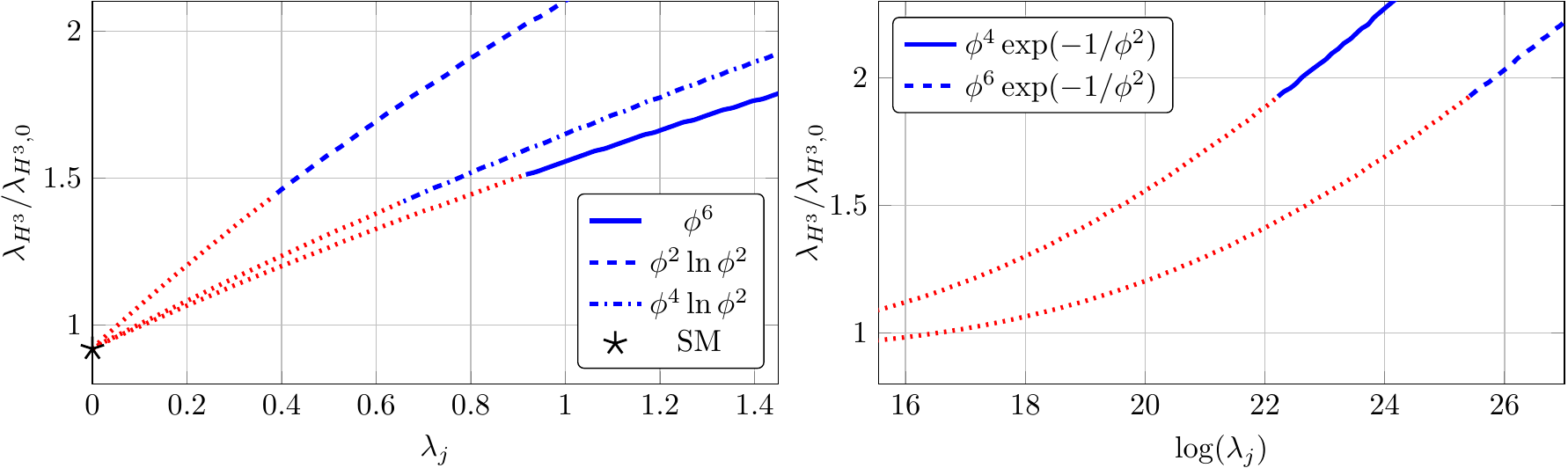}
 \caption{Modification of the self-coupling
   $\lambda_{H^3}/\lambda_{H^3,0}$ as a function of the
   coefficients $\lambda_j$ from the different UV potentials given in
   Eq.\eqref{eq:pots}.  Blue lines represent first-order phase
   transitions and red dotted lines second-order phase transitions. The cutoff
   is $\Lambda = 2$~TeV.}
 \label{fig:Higgs-self-coeff}
\end{figure}

Finally, Fig.~\ref{fig:Higgs-self-coeff} explicitly shows the connection
between the strength of the observable effect at LHC scales, measured
by $\lambda_{H^3}/\lambda_{H^3,0}$ and the size of the new physics
contribution $\Delta V$ at the microscopic scale $\Lambda$, measured
by the value of the dimensionless coefficients $\lambda_j$. The nature
of the electroweak phase transition is encoded in the coloring of the lines.
The onset of the first-order phase transition is at values that can 
also be read off from Fig.~\ref{fig:Higgs-self}: for logarithmic modifications
we find the lowest value of $\lambda_{H^3}/\lambda_{H^3,0} \approx 1.4$,
for the $\phi^6$ modification $\lambda_{H^3}/\lambda_{H^3,0} \approx 1.5$,
and for exponential modifications $\lambda_{H^3}/\lambda_{H^3,0} \approx 1.9$.
This size of all modifications can be probed in the high-luminosity run at the LHC.
Importantly, the Higgs self-couplings grow continuously as a function of $\lambda_j$
while $\phi_c/T_c$ remains zero till the onset of the first-order phase transition
and only then starts to grow continuously.

\section{Outlook}
\label{sec:outlook}

Higgs pair production or the measurement of the Higgs self coupling is
an extraordinarily interesting LHC analysis. We find that it is well
motivated by modified Higgs potentials which allow for a strong
first-order electroweak phase transition and hence an explanation of
the observed matter vs anti-matter asymmetry. 
We have studied a wide range of such modifications to
the Higgs potential,
especially potentials that cannot be
expanded as an effective field theory. We used the functional
renormalization group to describe the dependence on the field value
$\phi$ and on the temperature $T$. For all classes of potentials considered
here, there exists an appropriate choice of model parameters, for which 
the phase transition is of first order and sufficiently strong, 
$\phi_c/T_c \gtrsim 1$.\medskip

Our numerical analysis indicates that the requirement
$\phi_c/T_c=1$ corresponds to a critical scale of the order of 10~TeV
for all our potentials, where the potentials become strongly coupled.
Below this scale we can rely on our assumed potentials to describe LHC
signals. We then found that a strong first-order phase transition
universally predicts an enhancement of the Higgs self-couplings
$\lambda_{H^3} \gtrsim 1.5 \lambda_{H^3,0}$ and $\lambda_{H^4} \gtrsim
4 \lambda_{H^4,0}$.  Extending earlier studies, we systematically
established this connection between a first-order transition and a
measurable deviation of the Higgs self couplings, employing a method
that can describe systems with multiple physical scales in a
controlled manner.  While it might be possible that a new physics
model features a strong first-order transition with all effects on
$\lambda_{H^{3/4}}$ canceling accidentally~\cite{Noble}, none of our
examples falls into this class. We conclude that a measurement of 
the Higgs self-couplings at the LHC indeed serves as an indirect probe of
a first-order phase transition and thus of electroweak
baryogenesis in generic setups.

On the other hand, we observed that it is possible to obtain large 
deviations in the Higgs self-interactions for our class of non-perturbative
potentials without the condition $\phi_c/T_c\geq 1$ being
fulfilled.  For example with an exponential modification of the Higgs
potential the physical Higgs self-coupling reaches $\lambda_{H^3}
\approx 1.9 \lambda_{H^3,0}$ already significantly below $\phi_c/T_c=1$.
On the theoretical side, a quantitative upgrade 
of our analysis includes, but is not limited to, a full treatment 
of the weak gauge sector as well as improvements in our treatment 
of the Yukawa sector, which might result in quantitative changes 
of the order of 10 \%, cf.~\cite{Gies:2017zwf}.
An as precise as possible measurement of the triple-Higgs interaction is
clearly desirable. 
For instance a $20\%$ measurement of a relatively
small modification of $\lambda_{H^3}/\lambda_{H^3,0}$ could exclude
such exponential potentials as sources of electroweak baryogenesis.
Such an actual measurement could therefore provide valuable hints
guiding theoretical studies of interesting extended Higgs models.

\begin{center} \textbf{Acknowledgments} \end{center}

We thank R.~Sondenheimer for insightful discussions.  MR acknowledges funding
from IMPRS-PTFS and is grateful to the DFG research training group GRK 1523 at 
TPI Jena for hospitality.  JMP is supported by the Helmholtz Alliance HA216/EMMI
and by ERC-AdG-290623. AE is supported by an Emmy-Noether grant of the 
DFG under Ei-1037/1 and an Emmy-Noether visiting fellowship at the 
Perimeter Institute for Theoretical Physics. This work is part of and supported by the DFG
Collaborative Research Centre "SFB 1225 (ISOQUANT)".

\appendix
\section{Flow equations}
\label{app:der-flows}

The set of couplings in our setup consists of the $SU(3)$
coupling $g_3$, a fiducial coupling $g_F$ that simulates the $SU(2)$ and the
$U(1)$ sector, the top-Yukawa coupling $y_t$, and the full Higgs
potential $V(\phi)$ \cite{our_paper}. For the $SU(3)$ coupling it suffices to consider
one-loop running, since higher-order or threshold corrections have little
impact on the phase transition.  The one-loop beta function is
given by
\begin{align}
 \beta_{g_3} = -\frac{g_3^3}{(4 \pi)^2} \left(11 - \frac{2}{3} n_f\right) \,,
\end{align}
with $n_f=6$.  We fix the $SU(3)$ coupling through
$g_3(1\,\text{TeV})=1.06$, so the scale-dependent $SU(3)$ coupling
is known analytically. We approximate its temperature dependence by 
replacing $k \rightarrow \sqrt{k^2+\pi\, T^2}$,
\begin{align}
 g_3(k, T) = \left( \frac7{8 \pi ^2} \ln \frac{\sqrt{k^2+\pi T^2}}{1\,\text{TeV}} 
	  +\frac{1}{1.06^2}\right)^{-1/2} \,.
\end{align}
The logarithmic running of the $U(1)$ and $SU(2)$ couplings is
sufficiently slow to be negligible for our
purpose~\cite{our_paper}. We model it as a fiducial coupling $g_F$
that is a constant as a function of the RG scale and thus also a
constant as a function of the temperature.  At finite temperature,
this simplified treatment must be ameliorated by a thermal mass
generated by fluctuations from the electroweak sector.  According to
the high-$T$ expansion of the one-loop thermal potential it is given
by
\begin{align}
 V_{\text{thermal mass}} (\phi, T ) = \frac{1}{16}\left( 3g^2 + {g'}^{2} \right)
 \; \frac{T^2 \phi^2}{2} \,,
\end{align}
where $g=0.65$ and $g'=0.36$ are the $SU(2)$ and $U(1)$ gauge
couplings, respectively.\bigskip

To derive beta functions for the Higgs potential and the top-Yukawa
coupling we introduce the renormalized dimensionless field $\rho$ and
the dimensionless potential $u$
\begin{align}
 \rho = \frac{\phi^2}{2 k^2 Z_\phi}
 \qqqquad
 u(\rho) = \frac{V(\phi(\rho))}{k^4} \,.
\end{align}
The wave function renormalizations of the fields appear in the beta
functions only via their anomalous dimension
\begin{align}
 \eta_\phi = - \frac{\mathrm d \log Z_\phi }{\mathrm d \log k} 
 \qqqquad 
 \eta_\psi = - \frac{\mathrm d \log Z_\psi }{\mathrm d \log k} \,.
\end{align}
Written in terms of threshold functions, the beta function for the top
Yukawa coupling agrees with that from 
Refs.~\cite{our_paper,Gies:2013fua}, see, \eg~Eq.(C8) of
Ref.~\cite{our_paper}. However, we use a spatial regulator as
described below and temperature-dependent threshold functions. The
spatial regulator changes some prefactors, which is compensated by the
different definition of the threshold functions. The beta function is
given by
\begin{align} \label{eq:flow-eq-yuk} \frac{\mathrm d y_t^2}{\mathrm d
    \log k} ={}& y_t^2 \left(\eta_\phi+2 \eta_\psi\right) -
  \frac{y_t^4}{\pi^2} \left(3 \kappa u''(\kappa) + 2 \kappa^2
    u^{(3)}(\kappa)\right) l_{1,2}^{(FB)4}\left(\kappa
    y_t^2,u'(\kappa) + 2 \kappa
    u''(\kappa);\eta_\psi,\eta_\phi;T\right)
                 \notag \\
               &+ \frac{y_t^4}{2\pi^2}
                 \left(l_{1,1}^{(FB)4}\left(\kappa y_t^2,u'(\kappa) +
                 2 \kappa u''(\kappa);\eta_\psi,\eta_\phi;T\right)
                 \right. \notag \\
               &\left.\hspace{2cm}-2 \kappa y_t^2
                 \,l_{2,1}^{(FB)4}\left(\kappa y_t^2,u'(\kappa) + 2
                 \kappa u''(\kappa);\eta_\psi,\eta_\phi;T\right)
                 \right)
                 \notag \\
               &+\frac{3}{\pi^2} \frac{\left(N_c^2-1\right)}{2N_c}
                 g_3^2 y_t^2 \left(2 \kappa y_t^2\,
                 l_{2,1}^{(FB)4}(\kappa y_t^2,0;\eta_\psi,\eta_A;T)
                 - l_{1,1}^{(FB)4}(\kappa y_t^2,0;\eta_\psi,\eta_A;T)\right) \notag\\
               &-\frac{c_y g_F^2 y_t^2}{16 \pi ^2 \left(1+
                 \left(\frac{80}{246}\right)^2\kappa\right)}\,,
\end{align}
where $c_y=97/30$ and $N_c=3$. It depends on the position of the
renormalized dimensionless minimum $\kappa$ of the potential, the
anomalous dimensions of the fields, as well as on regulator-dependent
threshold functions specified below. Here, we have employed the same
projection scheme onto the Yukawa flow as in \cite{our_paper} for
reasons of comparison. In principle, there exists an improved scheme
\cite{Pawlowski:2014zaa} more adequately capturing higher-order
contributions to the Yukawa flow for the present model
\cite{Gies:2017zwf}, possibly improving the fixing of initial
conditions on the 5\% level. In either case, working in the symmetric
regime with $\kappa=0$ and neglecting the additional $\eta$ dependence
in the threshold functions reproduces the universal one-loop beta
functions, as it should.

The beta function for the Higgs potential at vanishing temperature has
been computed in Ref.~\cite{our_paper,Gies:2013fua}, see, \eg~Eq.(E1)
of Ref.~\cite{our_paper}. As for the beta function of the Yukawa
coupling, the present finite temperature beta function for the Higgs potential 
agrees with the $T=0$ one in terms of the threshold functions
\begin{align} \label{eq:flow-eq-pot} \frac{\mathrm d u (\rho)}{\mathrm
    d \log k}={}& -4 u(\rho) +(2+\eta_\phi) \rho\, u'(\rho) \notag
  \\&+\frac{1}{4\pi^2}\left( l_0^{(B)4}\left(u'(\rho)+2 \rho\,
      u''(\rho);\eta_\phi;T\right) - 4 N_c l_0^{(F)4}(y_t^2\rho
    ;\eta_\psi;T)\right) +\frac {c_l}{2 \pi ^2 \left(1 + \frac{g_F^2
        \rho}{2}\right)} \,,
\end{align}
where $c_l=9/16$ and again $N_c=3$.

Finally, we need expressions for the anomalous dimensions of the Higgs
field and the top-quark: the first two terms in
Eq.\eqref{eq:flow-eq-yuk} are integral parts of the universal one-loop
contribution.  In terms of the threshold functions the anomalous
dimension of the top quark agrees with the $T=0$ one in Eq.(C8) of
Ref.~\cite{our_paper}, and the anomalous dimension of the scalar field
has the same form as in Eq.(16) of Ref.~\cite{Gies:2013fua}. With the
thermal threshold functions of the present work this means
\begin{align} \label{eq:anom-dim} \eta_\phi ={}& \frac{2}{3\pi^2} N_c
    y_t^2 \left(m_4^{(F)4}(\kappa y_t^2;\eta_\psi;T)
    - \kappa y_t^2 m_2^{(F)4}(\kappa y_t^2;\eta_\psi;T) \right) \notag\\  
  &+ \frac{1}{3\pi^2} \kappa \left(3 u''(\kappa) 
    + 2 \kappa u^{(3)}(\kappa)\right)^2 m_4^{(B)4}\left(u'(\kappa) 
    + 2 \kappa u''(\kappa);\eta_\phi;T\right) \,,\notag\\
\eta_\psi ={}& \frac{1}{6\pi^2} y_t^2 \, 
    m_{1,2}^{(FB)4}\left(\kappa y_t^2,u'(\kappa) + 2 \kappa u''(\kappa);\eta_\phi;T\right)\notag\\
  &+ \frac{1}{2\pi^2} \frac{\left(N_c^2-1\right)}{ 2N_c} g_3^2
    \left(m_{1,2}^{(FB)4}(\kappa y_t^2,0;\eta_\psi,0;T)
    -\tilde{m}_{1,1}^{(FB)4}(\kappa y_t^2,0;\eta_\psi,0;T)\right) \,.
\end{align}
The beta functions found above are expressed in terms of
regulator-dependent and temperature-dependent threshold functions.
Here we provide explicit analytic results for these threshold
functions for one specific regulator. The analyticity of the
threshold function is rooted in the use of a Litim-type
regulator~\cite{flat-reg} that only regularizes the spatial
momenta. The dimensionless bosonic and fermionic propagators are
regularized as
\begin{align}
 G_\phi(\omega_n^2,{\vec p\,}^2 , m_\phi^2) = 
 \left(\omega_n^2 + {\vec p\,}^2/k^2 (1+r_B({\vec p\,}^2/k^2))+m_\phi^2\right)^{-1}\,,\notag\\
 G_\psi(\nu_n^2,{\vec p\,}^2 , m_\psi^2) = 
 \left(\nu_n^2 + {\vec p\,}^2/k^2 (1+r_F({\vec p\,}^2/k^2))+m_\psi^2\right)^{-1}\,,
\end{align}
with the bosonic Matsubara frequency $\omega_n=2\pi n T/k$ and the
fermionic Matsubara frequency $\nu_n = 2\pi (n+\frac12)T/k$.  Note
that $m_\phi$ and $m_\psi$ are dimensionless mass-like arguments.  The
bosonic and fermionic regulator shape functions read~\cite{flat-reg}
\begin{align}
 r_B(x) = \left(x^{-1}-1\right)\Theta(1-x)\,,\qqquad
 r_F(x) = \left(x^{-1/2}-1\right)\Theta(1-x)\,,
 \label{eq:flat-reg}
\end{align}
where  $x={\vec p\,}^2/k^2$.
In the following, we express the threshold functions in terms of
the bosonic and fermionic distribution functions,
\begin{align}
  n_{F,B}(m_{\psi,\phi}^2,T) = \left(\exp\left(\frac{k}{T} \sqrt{1+m_{\psi,\phi}^2}\right)\mp 1\right)^{-1}\; .
\end{align}
The set of threshold functions we need in our calculation includes
\begin{align}
 l_0^{(B)d}(m_\phi^2;\eta_\phi;T) ={}& \frac{2}{d-1} \left(1-\frac{\eta_\phi}{d+1}\right) \mathcal{B}_{(1)} (m_\phi^2;T) \,,\notag \\
 l_0^{(F)d}(m_\psi^2;\eta_\psi;T) ={}& \frac{2}{d-1} \left(1-\frac{\eta_\psi}{d}\right) \mathcal{F}_{(1)} (m_\psi^2;T) \,,\notag \\
 l_{n,m}^{(FB)d}(m_\psi^2,m_\phi^2;\eta_\psi,\eta_\phi;T) ={}& \frac{2}{d-1} 
 \left(n \left( 1 - \frac{\eta_\psi}d \right) \mathcal{FB}_{(n+1,m)}(m_\psi^2,m_\phi^2;T)\right.  
 \notag \\&\left.\hspace{1.7cm}
 + m \left( 1 - \frac{\eta_\phi}{d+1} \right) \mathcal{FB}_{(n,m+1)}(m_\psi^2,m_\phi^2;T) \right) \,,\notag 
 \end{align}
 \begin{align}
 m_4^{(B)d}\left(m_\phi^2;\eta_\phi;T\right) &= \mathcal{B}_{(4)}(m_\phi^2;T) \,,\notag \\
 m_{2}^{(F)d}(m_\psi^2;T) &= \mathcal{F}_{(4)}(m_\psi^2;T)  \,,\notag \\
 m_{4}^{(F)d}(m_\psi^2;\eta_\psi;T) &= \mathcal{F}_{(4)}(m_\psi^2;T) 
 + \frac{1- \eta_\psi}{d-3} \mathcal{F}_{(3)}(m_\psi^2;T) 
 - \frac12\left(\frac{1- \eta_\psi}{ d - 3}  + \frac{1}{2} \right) \mathcal{F}_{(2)}(m_\psi^2;T) \,,\notag \\
 m_{1,2}^{(FB)d}(m_\psi^2,m_\phi^2;\eta_\psi,\eta_\phi;T) &= 
 \left(1- \frac{\eta_\phi}d\right) \mathcal{FB}_{(1,2)}(m_\psi^2,m_\phi^2;T) \,,\notag \\
 \tilde{m}_{1,1}^{(FB)d}(m_\psi^2,m_\phi^2;\eta_\psi,\eta_\phi;T) &= 
 \frac{2}{d-2} \left( \left(1- \frac{\eta_\phi}d\right) \mathcal{FB}_{(1,2)}(m_\psi^2,m_\phi^2;T) \right. \notag \\
 &\qqquad+  \left(1- \frac{\eta_\psi}{d-1}\right) \mathcal{FB}_{(2,1)}(m_\psi^2,m_\phi^2;T) \notag \\
 &\qqquad\left.- \frac{1}{2} \left(1- \frac{\eta_\psi}{d-1}\right) \mathcal{FB}_{(1,1)}(m_\psi^2,m_\phi^2;T) \right) \,.
\end{align}
All threshold functions are expressed in terms of
\begin{align}
  \mathcal{F}_{(1)} (m_\psi^2;T) &= \frac{T}{k} \Mats G_\psi (\nu_n,m_\psi^2) \,,\notag \\
 \mathcal{B}_{(1)} (m_\phi^2;T) &= \frac{T}{k} \Mats G_\phi (\omega_n,m_\phi^2) \,,\notag \\
 \mathcal{FB}_{(1,1)}(m_\psi^2,m_\phi^2;T) &= \frac{T}{k} 
 \Mats G_\psi(\nu_n,m_\psi^2) G_\phi(\omega_n,m_\phi^2)  \,.
\end{align}
At finite temperature for the flat regulators in Eq.\eqref{eq:flat-reg} they
are given by
\begin{align}
 \mathcal{F}_{(1)} (m_\psi^2;T) ={}& \frac{1}{\sqrt{ 1 + m_\psi^2}} \left( \012 -  n_F ( m_\psi^2 , T ) \right)\,,\notag \\
 \mathcal{B}_{(1)} (m_\phi^2;T) ={}& \frac{1}{\sqrt{ 1 + m_\phi^2}} \left( \012 +  n_B ( m_\phi^2 , T ) \right)\,,\notag \\
 \mathcal{FB}_{(1,1)}(m_\psi^2,m_\phi^2;T) ={}& 
      \Bigg[ \frac{\frac{1}{2} + n_B(m_\phi^2 , T) }{2\sqrt{1+m_\phi^2}} 
      \left( \left(m_\psi^2 +1 - \left(i\pi T /k + \sqrt{1+m_\phi^2} \right)^2\right)^{-1}\right.\notag \\ 
      &\hspace{3.5cm}\left.+\left(m_\psi^2 +1 - 
      \left(i\pi T /k - \sqrt{1+m_\phi^2} \right)^2\right)^{-1} \right)  \notag \\
      &\,\,+\frac{ \frac{1}{2} - n_F( m_\psi^2 , T )}{2\sqrt{1+m_\psi^2}}
      \left( \left(m_\phi^2 +1 - \left(i\pi T /k + \sqrt{1+m_\psi^2} \right)^2\right)^{-1}\right.\notag \\ 
      &\hspace{3.5cm} +\left. \left(m_\phi^2 +1 - 
      \left(i\pi T /k - \sqrt{1+m_\psi^2} \right)^2\right)^{-1} \right) \Bigg]\,.
\end{align}
They obey the relations
\begin{alignat}{9} 
 \frac{\partial\mathcal{F}_{(n)} }{\partial m_\psi^2}   &= -n  \mathcal{F}_{(n+1)}\,, \qqqquad & 
 \frac{\partial\mathcal{B}_{(n)} }{\partial m_\phi^2}   &= -n  \mathcal{B}_{(n+1)} \,,\notag \\[1ex]
 \frac{\partial\mathcal{FB}_{(m,n)}}{\partial m_\psi^2}  &= -m \mathcal{FB}_{(m+1,n)}\,, \qqqquad & 
 \frac{\partial\mathcal{FB}_{(m,n)}}{\partial m_\phi^2}  &= -n \mathcal{FB}_{(m,n+1)}\,.
\end{alignat}
The notation and the threshold functions agree with
Ref.~\cite{Pawlowski:2014zaa}.  Note, that the $T\to0$ limit of the
threshold functions does not agree with the ones given in
Ref.~\cite{Gies:2013fua}, since we use a spatial regulator while
Ref.~\cite{Gies:2013fua} uses a covariant regulator.  This concludes
the list of threshold functions and relations necessary in order to
numerically evaluate the previously given beta functions.

\section{Grid approach and benchmarking}
\label{app:grid-code}

We solve the functional differential equation for the
Higgs potential, Eq.\eqref{eq:flow-eq-pot}, using a grid code. This means that the
potential $u(\rho)$ and its derivative $u^\prime(\rho)$ are
discretized on a grid in the field invariant $\rho$. 
The discretization converts the partial differential equation for $u(\rho)$ into a large
set of coupled ordinary differential equations.  The grid code has to manage a
numerical integration from $k=\Lambda$, where we initialize the flow,
down to $k=k_\text{IR} \approx 100$~GeV. At
this IR value all physical relevant quantities are frozen out and
only convexity-generating processes take place.

The grid code also has to cover a large range of values in the scalar
field $ 0 \leq \phi \leq c \Lambda$, where we typically choose $c =
\mathcal{O}(1\dots 10)$. To resolve both, large field values and the
minimum of the potential at small field values, we employ an
exponential distribution of the grid points $\rho_i=\phi_i^2/2$ with
$i\in{0,\ldots,N-1}$ according to
\begin{align}
 \rho_i = \rho_\text{a} + \frac{\exp\left(\frac{i}{c_\text{grid}}\right)-1}{\exp\left(\frac{(N-1)}{c_\text{grid}}\right)}\rho_\text{b}\,,
\end{align}
where $N$ is the number of grid points, $c_\text{grid}$ a grid
parameter that governs the distributions of the grid points, and
$\rho_\text{a}$ and $\rho_\text{b}$ the smallest and largest included
field value, respectively.

We introduce a grid for the potential $u(\rho_i)$ as well as for the
derivative of the potential $u'(\rho_i)$, and we match the second and
third derivative of the potential in between the grid
points~\cite{grid-code}. This is augmented by a differential equation
for the top-Yukawa coupling, while the $SU(3)$ coupling is already
integrated out and the fiducial coupling for $SU(2)$ and $U(1)$
remains constant. Consequently, we obtain a system of $2N + 1$ coupled
differential equations for a grid consisting of $N$~points, which is
solved with an iterative Runge-Kutta-Fehlberg method with an adaptive
step size.\bigskip

At the IR scale and at vanishing temperature, we match the output of
the grid code with the physically known observables, see
Eq.\eqref{eq:ir_data}.  This is implemented on the level of the
variables of the grid code and in particular we demand that the errors
fulfill $\Delta \rho_\text{min} \leq 20$~GeV$^2$, $\Delta \lambda_4
\leq0.002$ and $\Delta y_t \leq 0.0014$.  Expressed in the quantities
of Eq.\eqref{eq:ir_data} these errors correspond to $\Delta v \leq
0.08$~GeV, $\Delta m_H \leq 0.28$~GeV, and $\Delta m_t \leq 0.23$~GeV.
It is important to determine the vacuum expectation value more
precisely since its error directly influences the error on the Higgs
and the top mass.

To achieve this precision we tune the parameters $\mu$,
$\lambda_4$ and $y_t$ at the UV scale, which is done by a secant
method in $\mu$ and a two-dimensional bisection method in $\lambda_4$
and $y_t$ .  The grid code might exhibit other systematic errors and
in particular the measurement of the Higgs mass is challenging since
it is related to the second derivative of the potential.  Hence we
conservatively estimate the total accuracy of the IR values with
\begin{align}
 \Delta v \leq 0.2 \ \gev\,, \qqquad \Delta m_H \leq 1.5 \ \gev\,,  \qqquad \Delta m_t \leq 0.5 \ \gev \,.
\end{align}
The tuning process is performed at vanishing temperature and the
tuned initial values are subsequently used as initial values for
all finite-temperature computations.  For each temperature we
initialize the flow in this way and determine the position of the
minimum at the IR scale $k_\text{IR}$.  The critical temperature is
obtained with a bisection method where we demand an accuracy of
$\Delta T_c \leq 0.2$~MeV.  This high accuracy is necessary for a
precise value of $\phi_c$, which is in turn given by the position of
the minimum at the temperature just below $T_c$.  From the grid code, it
is difficult to get a clear signature
distinguishing between second-order phase transitions and weak
first-order phase transitions.  
Within our numerical accuracy, a reliable distinguishing signature is not available
for
$\phi_c\lesssim20$~GeV.  For finite temperature computations we 
slightly increase the number of grid points, since the exponential
functions in the bosonic and fermionic distribution functions make
these computations technically more challenging.\bigskip

\begin{figure}[t]
 \includegraphics[width=\textwidth]{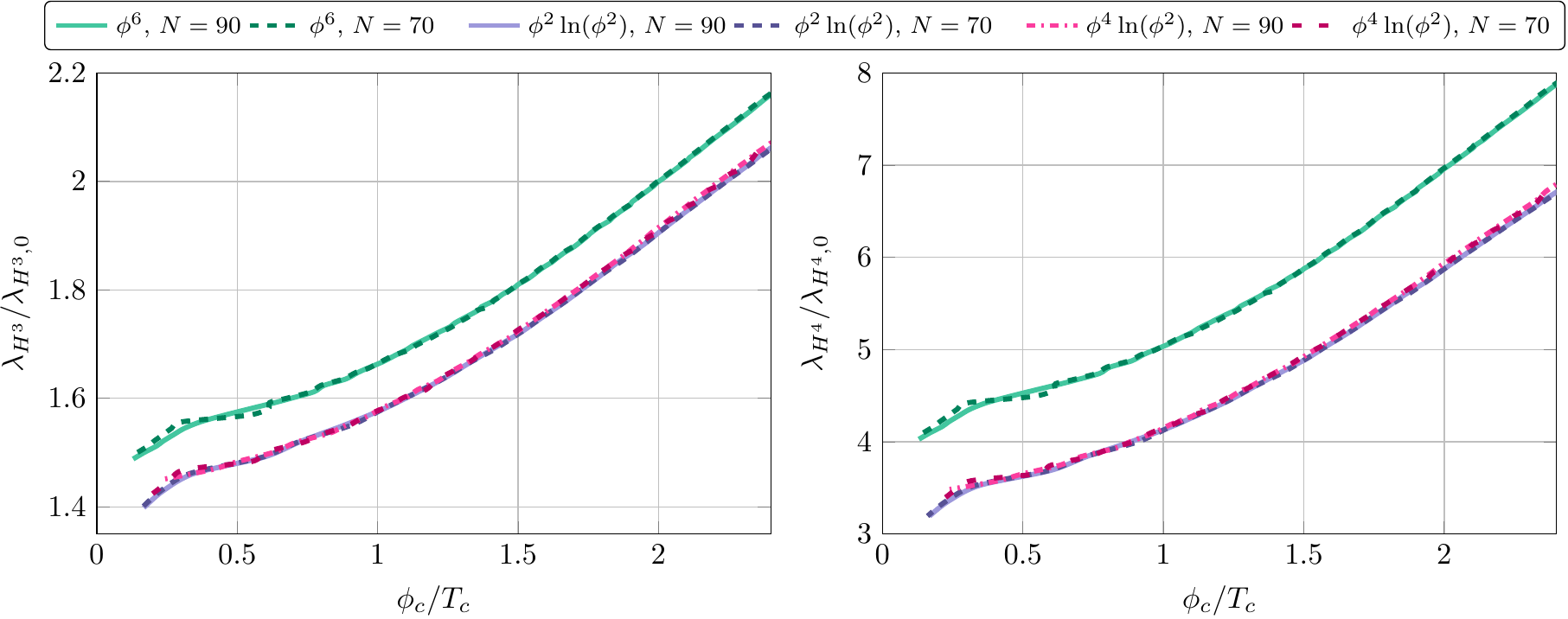}
 \caption{Modification of the self-couplings
   $\lambda_{H^3}/\lambda_{H^3,0}$ (left) and
   $\lambda_{H^4}/\lambda_{H^4,0}$ (right) as a function of
   $\phi_c/T_c$ for polynomial and logarithmic modifications of the UV
   potentials, cf.~Eq.\eqref{eq:pots}.  We compare results for $N=70$
   and $N=90$ grid points.}
 \label{fig:convergence}
\end{figure}

\begin{figure}[b!]
 \includegraphics[width=\textwidth]{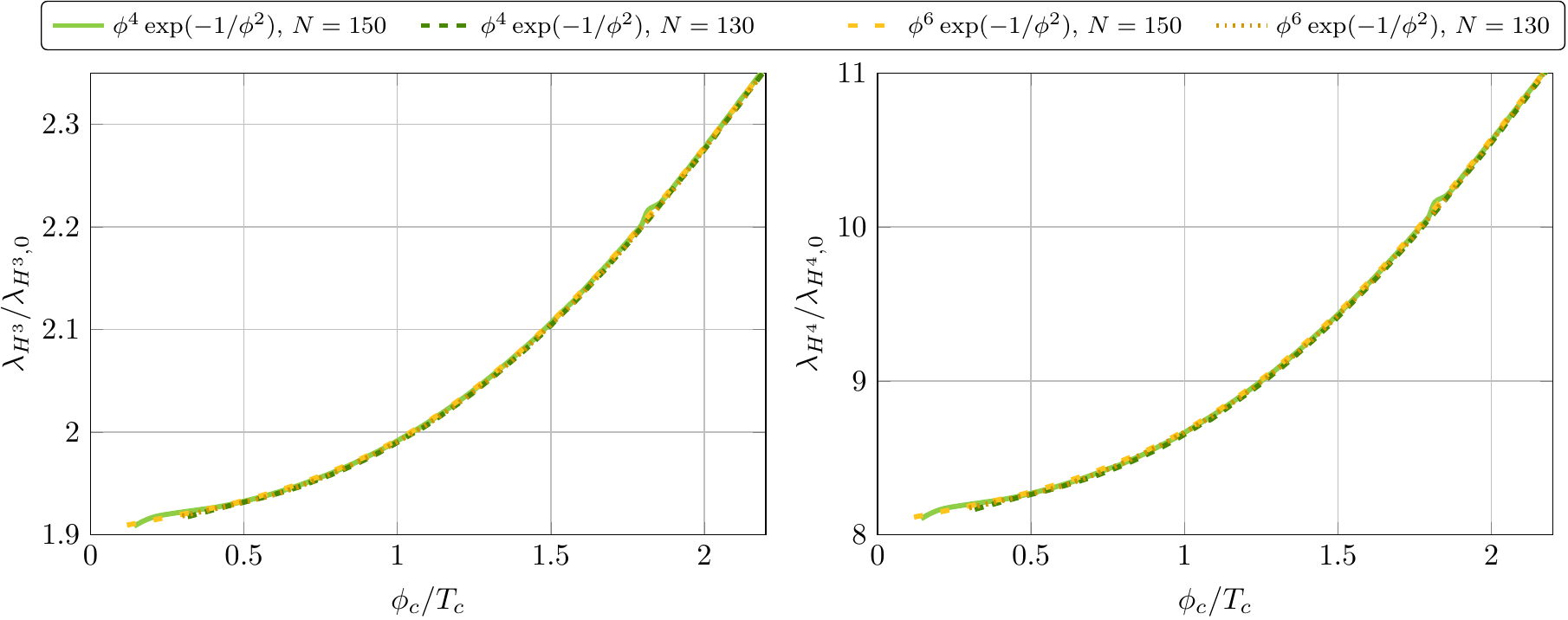}
 \caption{Modification of the self-couplings
   $\lambda_{H^3}/\lambda_{H^3,0}$ (left) and
   $\lambda_{H^4}/\lambda_{H^4,0}$ (right) as a function of
   $\phi_c/T_c$ for exponential modifications of the UV potentials,
   cf.~Eq.\eqref{eq:pots}.  We compare results for $N=130$ and
   $N=150$ grid points.}
 \label{fig:convergence-exp}
\end{figure}

We test our numerical results by first comparing the observables for two
different numbers of grid points.  The necessary number varies with
our choice of cutoff and the modification of the Higgs potential.  For
example, more grid points are necessary for the exponential
modifications of the potential.  For polynomial and logarithmic
modifications and a cutoff $\Lambda=2$~TeV, we use typically $N=90$
grid points, while for exponential modifications with the same cutoff
we use $N=150$ grid points.
In Fig.~\ref{fig:convergence} we display results for polynomial and
logarithmic modifications. In
particular we show the correlation between the strength of
first-order phase transition and the Higgs-self couplings.  In
Fig.~\ref{fig:convergence-exp} we show the same correlation but for
exponential modifications and for $N=130$ and for $N=150$ grid points.
The results for $N=90$ and for $N=150$ are identical with those
displayed in Fig.~\ref{fig:Higgs-self}.

\begin{figure}[t]
 \includegraphics[width=.6\textwidth]{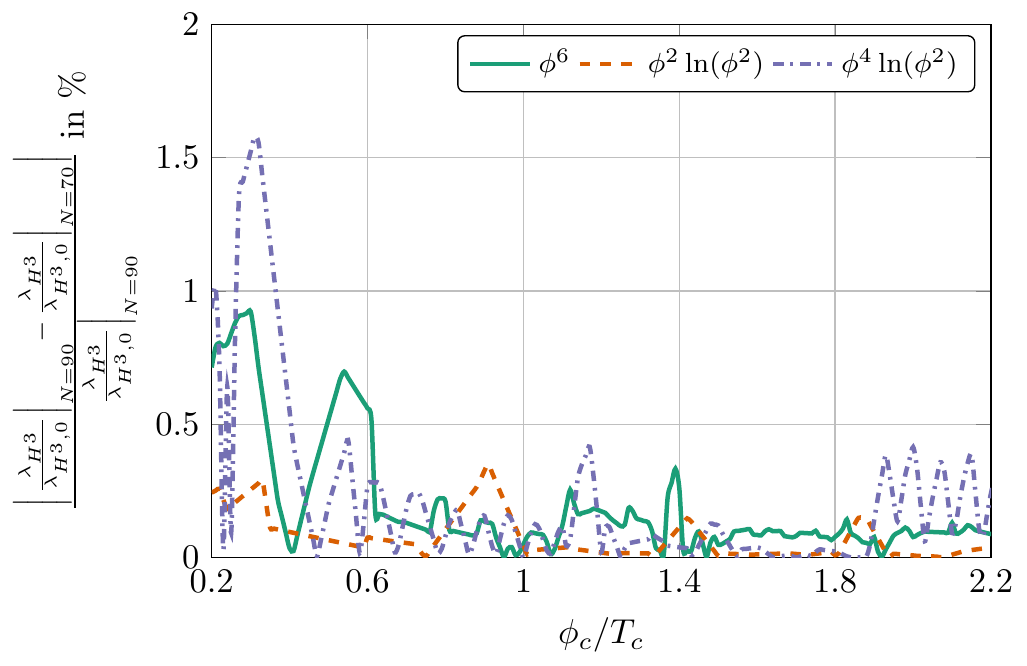}
 \caption{Relative change of $\lambda_{H^3}/\lambda_{H^3,0}$ with
   different numbers of grid points as a function of $\phi_c/T_c$ for
   polynomial and logarithmic modifications of the UV potentials,
   cf.~Eq.\eqref{eq:pots}.  In the regime of interest of $\phi_c/T_c
   \geq 1$, the relative difference between $N=70$ and $N=90$ is in
   the sub-percent regime, $\leq 0.5 \%$.}
 \label{fig:relative-error}
\end{figure}

To make our analysis more quantitative we also display the relative
change of the correlation for polynomial and logarithmic modifications
in Fig.~\ref{fig:relative-error}.  The results do not change
significantly when we increase the number of grid points.  In case of
polynomial and logarithmic modifications the amount of wiggles in the
region of a weak first-order phase transition, which originates from
numerical uncertainties, is further reduced.  In the region of a weak
first-order phase transition we have a relative change of less than
2\%, while in the region of a strong first-order phase transition we
have a relative change of less than 0.5\%.  This is sufficient for our
analysis, since we are only interested in the latter case.  In case of
the exponential modifications the change is hardly visible.  The
relative change is globally less than 0.02\%. These results illustrate
that our findings are indeed numerically stable.\bigskip

Finally, we can compare our functional renormalization group results
to other methods, for instance to the mean-field-like methods of
Ref.~\cite{christophe_geraldine}.  To perform a meaningful comparison,
we have to take into account the slightly different setup: while we
modify the microscopic potential, Ref.~\cite{christophe_geraldine}
implements the modifications directly at the level of the effective
potential. This means that in our setup a $\phi^6$ modification of the
microscopic potential generates finite higher-order modifications
through quantum fluctuations, which in the weak coupling regime are
similar to the one-loop determinant. These additional terms do not
appear in Ref.~\cite{christophe_geraldine}.

For our comparison we therefore adjust the parameter $\lambda_6$ such
that the $T=0$ effective potentials of both setups agree. Due to the
impact of quantum fluctuations, different values of $\Lambda$ require
slightly different initial conditions for $\lambda_6$ in our setup.
With a cutoff $\Lambda=1$~TeV it turns out that this is the case for
$\lambda_6\approx 0.21$, while for a cutoff $\Lambda=0.6$~TeV we find
$\lambda_6\approx 0.19$. The difference in values of $\lambda_6$ is
accounted for by the RG flow between the two choices of cutoff scale.
With these values we can then compare $T_c$ and $\phi_c/T_c$.  As
expected, we indeed find good qualitative agreement.  For instance,
for $\Lambda=0.6$~TeV we find $\phi_c/T_c=2.7$ and $T_c=83$~GeV vs
$\phi_c/T_c=2.8$ and $T_c=75$~GeV from
Ref.~\cite{christophe_geraldine}.  We emphasize that a more precise
agreement cannot be expected: the modification of the microscopic and
the effective Higgs potential are necessarily different, as our setup
accounts for quantum fluctuations, in particular affecting
$\lambda_6$ between the microscopic scale and the IR.


\end{document}